\documentclass[11pt, draftclsnofoot, journal, onecolumn, a4]{IEEEtran}
\usepackage{amsmath,cite,amsfonts,amssymb,psfrag}
\usepackage{graphicx}
\usepackage{rotating}
\usepackage{color}
\usepackage{epstopdf}

\newtheorem{theorem}{Theorem}

\newtheorem{remark}{Remark}
\newtheorem{definition}{Definition}

\newtheorem{lemma}{Lemma}

\newfont{\bbb}{msbm10 scaled 700}

\newfont{\bb}{msbm10 scaled 1100}

\newcommand{\PP}{\mbox{\bb P}}

\newcommand{\Wc}{{\cal W}}

\renewcommand{\arg}{{\hbox{arg}}}

\usepackage{graphicx,amssymb,amstext,amsmath,color}
\DeclareGraphicsExtensions{.eps,.pdf,.png,.jpg,.gif,.jpeg,.pstex}

\usepackage{hyperref}
\hypersetup{
    bookmarks=true,         
    unicode=false,          
    pdftoolbar=true,        
    pdfmenubar=true,        
    pdffitwindow=false,     
    pdfstartview={FitH},    
    pdfnewwindow=true,      
    colorlinks=true,       
    linkcolor=red,          
    citecolor=green,        
    filecolor=magenta,      
    urlcolor=cyan           
}

\begin{document}
\title{Secret key-based Identification and Authentication with a Privacy Constraint}
\author{\IEEEauthorblockN{Kittipong Kittichokechai and Giuseppe Caire}\\
\IEEEauthorblockA{Technische Universit\"{a}t Berlin}
\thanks{This work was partially supported by an Alexander von Humboldt Professorship Grant}
}

\maketitle

\begin{abstract}
We consider the problem of identification and authentication based on secret key generation from some user-generated source data (e.g., a biometric source). 
The goal is to reliably identify users pre-enrolled in a database as well as authenticate them based on the estimated secret key while preserving the 
privacy of the enrolled data and of the generated keys. We characterize the optimal tradeoff between the identification rate, 
the compression rate of the users' source data, information leakage rate, and secret key rate. In particular, we provide a coding strategy based on 
layered random binning which is shown to be optimal. In addition, we study a related secure identification/authentication 
problem where an adversary tries to deceive the system using its own data. 
Here the optimal tradeoff between the identification rate, compression rate, leakage rate, and exponent of the maximum false acceptance probability 
is provided.  The results reveal a close connection between the optimal secret key rate and the false acceptance exponent 
of the identification/authentication system.
\end{abstract}


\newpage

\section{Introduction}\label{sec:introduction}

\begin{figure}[ht]
\centerline{\includegraphics[width=13cm]{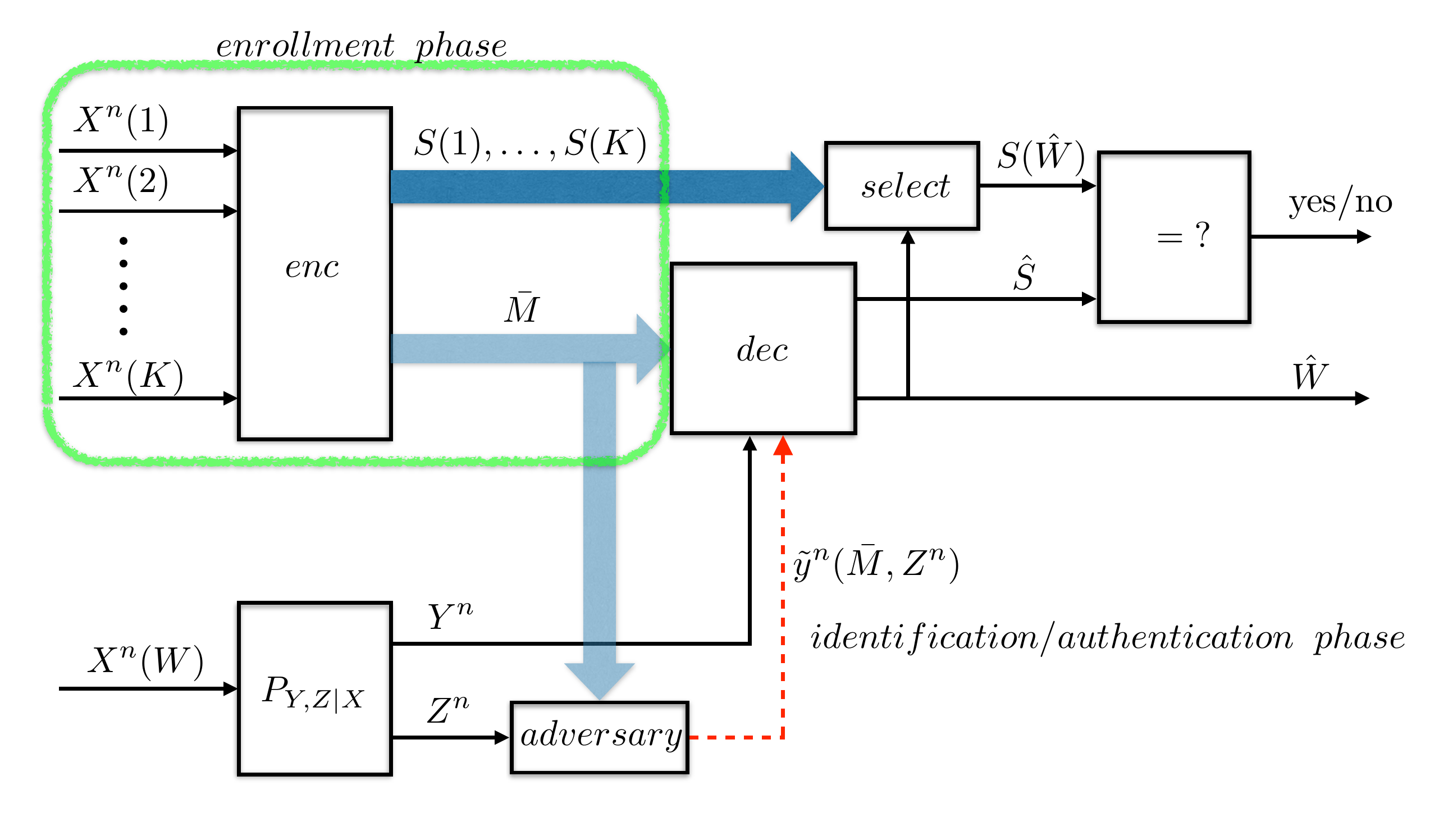}}
\caption{Identification and secret key-based authentication system with an adversary. The enrollment phase is contained in the green box.
The remaining part corresponds to the identification/authentication phase. The red dashed arrow corresponds to an active adversary which replaces
the original user measurement $Y^n$ with its own generated signal $\tilde{y}^n(\bar{M},Z^n)$ in order to gain access to the system.}
\label{fig:scheme}
\end{figure}

Consider an identification and authentication system with $K$ users (see Fig. \ref{fig:scheme}). 
In the enrollment phase, each user $w \in \{1,2,\ldots,K\}$ generates a source sequence $X^n(w)$ and provides it to the system. 
Such source sequences are compressed into $\bar{M}\triangleq \{M(w) : w = 1, \ldots, K\}$ and stored into a database. The compressed user source data
will be used as a reference for identification of the enrolled users.   At the same time, the system produces a set of secret keys $\{ S(w) : w = 1, \ldots, K\}$, also functions of the users' source sequences,  which will be used as a reference for authentication of the identified user.
In the identification/authentication phase, an a-priori unknown user $W$ provides a measurement $Y^n$ to the system. 
For example, this could be seen as a  noisy version of its enrolled source sequence $X^n(W)$. 
Based on the stored database $\bar{M}$ and measurement $Y^n$, the user is identified 
as $\hat{W}$. The system also produces an estimated key $\hat{S}$. 
The user is successfully identified and authenticated if  $(\hat{W},\hat{S}) = (W,S(W))$.

The system described above can be relevant in several applications including those involving access control, secure, and trustworthy communication. 
In database identification for access control applications, the system identifies an individual as an enrolled user and then grants the corresponding 
access based on authentication using the secret-key. In other words, the system first finds out which user in the database the individual corresponds to, 
and then verifies whether the individual is really the user he/she claims to be. 

One important class of access control applications is related to using biometric data such as fingerprint, iris scans, voice, face, 
and DNA sequences (see, e.g., \cite{RaneSurvey} and references therein). Unlike passwords, the biometric data inherently belong to the users and provide a convenient  and seemingly more secure way for identification/authentication. However, it is crucial that privacy of the enrolled data must be protected 
from any inference of an adversary. The privacy risk in this case is of potentially high impact since the biometric data is typically  tied to the person identity. 
If it is compromised, it cannot be reverted or changed easily, unlike in the case of a password.

In this work, we consider secret-key based identification and authentication problems in the presence of an 
adversary which is not part of the system but has full knowledge of the stored database data  $\bar{M}$ as well as to some ``on-line'' side information
$Z^n$, as shown in Fig.~\ref{fig:scheme}. We refer to $Z^n$ as on-line side information since it is statistically dependent 
on the actual user $W$ which is trying to be identified and authenticated. In contrast, the knowledge of $\bar{M}$ can be regarded as
``off-line'' side information for the adversary. Two closely related scenarios are studied:

\begin{itemize}

\item[1)]
The adversary is passive and is only interested in inferring the user source sequence. In this case, 
we wish to design a reliable identification/authentication system that achieves maximal identification 
rate and secret key rate (see definitions in Section \ref{sec:problem_setting_keyrate})  
while minimizing the the compression rate of the stored descriptions
and the information leakage of the enrolled source sequences. 
In general, there exists tension between these performance metrics.  For this scenario, our main contribution is a single-letter characterization of the 
optimal tradeoff region of the identification rate, compression rate, information leakage rate, and key rate for discrete memoryless sources.

\item[2)]
The adversary is active and tries to deceive the identification/authentication system by using its own sequence 
$Y^n = \tilde{y}^n(\bar{M},Z^n)$. We refer to the event where the legitimate user fails during the identification/authentication as a {\em false rejection}, 
and to the event where the system accepts the adversary as a {\em false acceptance}. In this case, 
we wish to design a secure identification/authentication system that achieves arbitrarily small false rejection probability 
with maximum identification rate and: i) minimizes the compression rate of each stored description, ii) minimizes the leakage rate of each enrolled 
sources, and iii) maximizes the error exponent of the maximum false acceptance probability (mFAP) (see definitions in Section \ref{sec:problem_setting_mFAP}). 
For this scenario, our main contribution  is a single-letter characterization of the optimal tradeoff between the identification rate, compression rate, information leakage rate, and mFAP exponent for discrete memoryless sources.
\end{itemize}

In order to motivate the role of key-based authentication to the possibly unfamiliar readers, we use the following naive everyday-life example.
Consider the front door of a building with an intercom with multiple buttons. Each button corresponds to an apartment in the building. 
An intruder may wish to gain access to the building by hitting at random a button, hoping that the people inside the corresponding 
apartment just open the door, by identifying the intruder as friend/family just because he/she hit their button. 
Instead, if the intercom is also equipped with a camera and a facial recognition software, the door will be opened only if 
the intruder face (properly projected into some features space) generates a hashing function value that matches with the 
key corresponding to that apartment.  Technically speaking, the optimal identification problem corresponds to $K$-ary hypothesis testing, which just 
provides the answer minimizing the average probability of wrong identification. However, the identified user needs also to be authenticated 
(in this case, by showing his/her face) in order to rightfully gain access to the system.

\subsection*{Related Work}

Authentication problems have been studied from an information theoretic perspective in several directions. 
Maurer in \cite{Maurer} considered the message authentication problem in connection with the hypothesis testing problem where the underlying message probability distributions of the legitimate user and adversary are assumed to be different. Martinian et al. \cite{Martinian} considered authentication with a distortion criterion. More recently, some works have considered authentication based on secret key generation \cite{AhlswedeCsiszar} which are closely 
related to fuzzy extractor \cite{Dodis}.  These include, for example, Lai et al., and Ignatenko and Willems \cite{LaiHoPoor},\cite{LaiHoPoor_b},\cite{IgnatenkoWillems_a}, \cite{IgnatenkoWillems},\cite{WillemsIgnatenko}, which focused on biometric authentication systems \cite{JuelsWattenberg} where privacy of the enrolled data is also taken into account. In \cite{Kittichokechai_b}, we considered a general case where the adversary has correlated 
side information and we provided a complete characterization of the fundamental tradeoff. Analysis of deception probability in authentication systems from an adversary's perspective was also considered in \cite{Kang}. Closely related to the secret key-based authentication problem with privacy constraint 
are the problems of source coding with privacy constraint, e.g.,  \cite{VillardPiantanida},\cite{Kittichokechai}, where the goals are to reconstruct 
the source reliably while preserving the privacy of the source or the reconstruction sequences from the inference of an eavesdropper.

By extending the single-user authentication problem to the identification/authentication problem in the multi-user case, another dimension 
is added into the problem, namely we also care for the identification rate. A database identification problem for biometric data was considered in \cite{O’Sullivan}, \cite{Willemsetal} where the noisy measurement of all user data are treated as a database and the maximum identification rate was characterized. 
Later, Tuncel \cite{Tuncel} considered the problem where the database is a compressed version of the user data and showed 
the optimal tradeoff between identification rate and compression rate. Recently, this was extended to include also a lossy reconstruction constraint 
at the decoder \cite{TuncelGunduz}. Ignatenko and Willems \cite{IgnatenkoWillems_iden} studied the problem of user identification 
together with secret key-based authentication under a privacy constraint, extending the secret-key based authentication problem 
to the multi-user setting.

\subsection*{Contribution and Organization}

In this work we extend the setting of \cite{IgnatenkoWillems_iden} to a more general case, including a compression 
rate constraint on the source description and allowing the adversary to have access to correlated side information.  
The setting of this paper can also be viewed as a multi-user extension of our previous work \cite{Kittichokechai_b}. 
Correlated side information at the adversary, as treated here, is of practical interest since it models scenarios where the adversary 
can have access to noisy version of the source data. 
In Section \ref{sectionII}, we study the secret key-based identification with a privacy constraint and provide a complete characterization of the identification-compression-leakage-key  rate region $\mathcal{R}_1$ for discrete memoryless sources. It is shown that the layered binning scheme with rate allocation between compression and identification only on the first-layer description is optimal. 
The result includes many other results as special cases, one of which is the compression-leakage-key rate region for secret key-based authentication problem in \cite{Kittichokechai_b}. Binary examples illustrating the derived tradeoffs are also provided.
In Section \ref{sectionIII}, we study a secure identification problem with a privacy constraint and provide a complete characterization of the identification-compression-leakage-mFAP exponent region $\mathcal{R}_2$ for discrete memoryless sources. 
Our results show that  the maximum key rate in $\mathcal{R}_1$ is equivalent to the maximum mFAP exponent in $\mathcal{R}_2$, 
revealing a connection between secret key rate and security of identification/authentication system.

\textit{Notation}: We denote discrete random variables, their corresponding realizations or deterministic values, and their alphabets by the upper case, lower case, and calligraphic letters, respectively. 
$X_{m}^{n}$ denotes the sequence $\{X_{m},\ldots,X_{n}\}$ when $m\leq n$, and the empty set otherwise. 
Also, we use the shorthand notation $X^{n}$ for $X_{1}^{n}$. The term $X^{n\setminus i}$ denotes the set $\{X_{1},\ldots,X_{i-1},X_{i+1},\ldots,X_{n}\}$. 
When a random variable $X$ is constant we write $X=\emptyset$.  
A length-$K$ vector of descriptions $(M(1),\ldots,M(K))$ is denoted by $\bar{M}$, where $\bar{M}^{\setminus W}$ is the vector $(M(1),\ldots,M(W-1),M(W+1),\ldots,M(K))$.  Cardinality of the set $\mathcal{X}$ is denoted by $|\mathcal{X}|$. We use $[1:N]$ to denote the index set $\{1,2,\dots,N\}$. 
Finally, we use $X-Y-Z$ to indicate that $(X,Y,Z)$ forms a Markov chain.  Other notations follow the standard ones in \cite{ElGamalKim}.

\section{Secret key-based Identification/Authentication with a Privacy Constraint} \label{sectionII}

\subsection{Problem Formulation}\label{sec:problem_setting_keyrate}
Let us consider a secret key-based identification and authentication system as shown in Fig. \ref{fig:scheme}. 
Source,  measurement and side information alphabets, $\mathcal{X}, \mathcal{Y}, \mathcal{Z}$ are finite sets.
The users' source sequences $X^{n}(w)$ for $w \in \Wc^{(n)} \triangleq [1:K]$ are independent across the users and have i.i.d. components distributed according to  some fixed source distribution $P_X$. 
In the enrollment phase, an {\em encoder} generates a description $M(w)$ and a secret key message $S(w)$ 
based on $X^n(w)$, for each $w \in \mathcal{W}^{(n)}$.  The descriptions are stored in a database for later identification and authentication.
In the identification/authentication phase, an arbitrary unknown user $W \in \mathcal{W}^{(n)}$, independent of the enrolled source sequences and stored database, presents itself to the system, and generates
measurement sequence $Y^n$ jointly distributed with $X^n(W)$. 
Based on $Y^n$ and the stored database $\bar{M}\triangleq(M(1),M(2),\ldots,M(K))$, a {\em decoder} identifies 
the observed user as $\hat{W}$ and generates an estimation of the key $\hat{S}$. 
The identification and authentication operation is successful if  $(\hat{W},\hat{S}) = (W,S(W))$.

We consider an adversary which has access to the whole database $\bar{M}$ and to a side information sequence 
$Z^n$ also jointly distributed with $X^n(W)$. The information leakage rate of user $W$ at the adversary is measured by the 
mutual information rate $I(X^n(W);\bar{M},Z^n)/n$.  
Similarly, the key leakage rate of user $W$ at the adversary is measured by the mutual information rate $I(S(W);\bar{M},Z^n)/n$. 

In this work we assume that $(X^n(W), Y^n,Z^n)$ are memoryless (with respect to the sequence index~$i$) 
with the $i$-th marginal joint distribution $P_{X,Y,Z} = P_X P_{Y,Z|X}$, where $P_{Y,Z|X}$ is a given transition probability distribution of 
a discrete memoryless broadcast channel (see Fig. \ref{fig:scheme}). In contrast, for all $w \neq W$, the triples
$(X^n(w), Y^n,Z^n)$ are memoryless with the $i$-th marginal distribution $P_X P_{Y,Z}$, where $P_{Y,Z}$ is the YZ-marginal distribution of $P_{X,Y,Z}$. 

We are interested in characterizing the optimal tradeoff between the identification rate, compression rate, information leakage rate, and secret key rate, defined as follows:

\begin{definition}\label{def:code_keyrate}
An $(|\mathcal{M}^{(n)}|,|\mathcal{W}^{(n)}|,|\mathcal{S}^{(n)}|,n)$-code for secret key-based identification and authentication with a privacy constraint consists of
\begin{itemize}
\item A set of stochastic encoders $F_{w}^{(n)}: w \in \mathcal{W}^{(n)}$ such that the $w$-th encoder takes $X^n(w)$ as input and generates $(M(w),S(w)) \in \mathcal{M}^{(n)} \times \mathcal{S}^{(n)}$ according to a conditional PMF $p(m(w),s(w)|x^n(w))$.
  \item A decoder $g_{\rm Id}^{(n)}: {\mathcal{M}^{(n)}}^K \times \mathcal{Y}^{n} \rightarrow \mathcal{W}^{(n)}$, 
  such that the identified user is $\hat{W} =  g_{\rm Id}^{(n)}(\bar{M}, Y^n)$. 
  \item A decoder $g_{\rm Au}^{(n)}: {\mathcal{M}^{(n)}}^K \times \mathcal{Y}^{n} \rightarrow \mathcal{S}^{(n)}$, 
  such that the estimated secret key is  $\hat{S} = g_{\rm Au}^{(n)}(\bar{M}, Y^n)$.  
\end{itemize}
\hfill $\lozenge$
\end{definition}

\begin{definition}  An identification-compression-leakage-keyrate tuple $(R_I,R_C,L,R_S) \in \mathbb{R}^{4}_{+}$ is said to be \emph{achievable} if, for any $\delta>0$ there exists a sequence of  $(|\mathcal{M}^{(n)}|,|\mathcal{W}^{(n)}|,|\mathcal{S}^{(n)}|,n)$-codes such that, for all sufficiently large $n$,
\begin{align}
\max_{W \in \Wc^{(n)}} \PP((\hat{W},\hat{S})\neq (W,S(W))) &\leq \delta, \label{eq:error_constraint}\\
\frac{1}{n}\log\big|\mathcal{W}^{(n)}\big| &\geq R_I-\delta,\label{eq:identification_rate_constraint}\\
\frac{1}{n}\log\big|\mathcal{M}^{(n)}\big| &\leq R_C+\delta,\label{eq:compression_rate_constraint}\\
\max_{W \in \mathcal{W}^{(n)}} \frac{1}{n}I(X^{n}(W);\bar{M},Z^n) &\leq L+\delta,\label{eq:leakage_constraint}\\
\max_{W \in \mathcal{W}^{(n)}}  \frac{1}{n}I(S(W);\bar{M},Z^n) &\leq \delta, \label{eq:key_leakage_constraint}\\
\min_{W \in \mathcal{W}^{(n)}}  \frac{1}{n}H(S(W)) &\geq R_S-\delta,\label{eq:key_rate_constraint}
\end{align}
The \emph{identification-compression-leakage-keyrate} region $\mathcal{R}_1$ is defined as the closure of all achievable tuples.
\hfill $\lozenge$
\end{definition}

\subsection{Results}

\begin{theorem}\label{theorem:region_keyrate}
The region $\mathcal{R}_1$ for the identification/authentication problem defined above 
is given by a set of  all tuples $(R_I,R_C,L,R_S)\in \mathbb{R}^{4}_{+}$ such that
\begin{align}
R_I &\leq I(Y;U),  \\
R_C &\geq R_I +I(X;V|Y),\label{eq:tradeoff_compression_identification}\\
L & \geq I(X;V,Y) - I(X;Y|U)+I(X;Z|U),\\
R_S &\leq I(V;Y|U)-I(V;Z|U),
\end{align}
for some $P_{X,Y,Z} P_{V|X} P_{U|V}$ with  $|\mathcal{U}| \leq |\mathcal{X}| +4, |\mathcal{V}| \leq (|\mathcal{X}|+4)(|\mathcal{X}|+2)$. 
\hfill $\square$
\end{theorem}

By standard time-sharing argument \cite{WillemsVDMeulen}, it is immediate to show that $\mathcal{R}_1$ is convex. 

Before giving the proof of Theorem \ref{theorem:region_keyrate}, some remarks are in order.

\begin{remark}[Layered random binning]
Binning usually helps to reduce the rate needed for compression. In the related identification problem \cite{TuncelGunduz} the authors showed that the binning scheme is optimal when an additional reconstruction constraint is included. As we shall see in the proof of  Theorem \ref{theorem:region_keyrate}, 
\emph{layered  binning} turns out to be also optimal in the presence of an information leakage constraint towards an adversary 
with access to correlated side information. Interestingly, we note that the obtained tradeoff between compression and identification 
rates in \eqref{eq:tradeoff_compression_identification} results from the rate allocation which is applied only on the first 
layered codeword.
\end{remark}

\begin{remark}[Special cases]\label{remark:special_cases} Theorem~\ref{theorem:region_keyrate} recovers results of several special cases in the literature.  \par
i) When there is only one user in the database, i.e., $|\mathcal{W}^{(n)}|=1$, the problem reduces to authentication with a privacy constraint studied in \cite{Kittichokechai_b} (see e.g., Fig. \ref{fig:authentication_keyrate}). It can also be viewed as an extension of the secret key agreement problem with one-way communication constraint \cite{CsiszarNarayan} to include an information leakage constraint. 
By setting $R_I=0$ in $\mathcal{R}_1$, we obtain the compression-leakage-keyrate region consisting of all tuples $(R_C,L,R_S)$ such that
\begin{align*}
R_C &\geq I(X;V|Y),\\
L & \geq I(X;V,Y) - I(X;Y|U)+I(X;Z|U),\\
R_S &\leq I(V;Y|U)-I(V;Z|U),
\end{align*}
for some joint distributions of the form $P_{X,Y,Z}P_{V|X}P_{U|V}$ with  
$|\mathcal{U}| \leq |\mathcal{X}| +3, |\mathcal{V}| \leq (|\mathcal{X}|+3)(|\mathcal{X}|+2)$.

\begin{figure}[ht]
\centerline{\includegraphics[width=13cm]{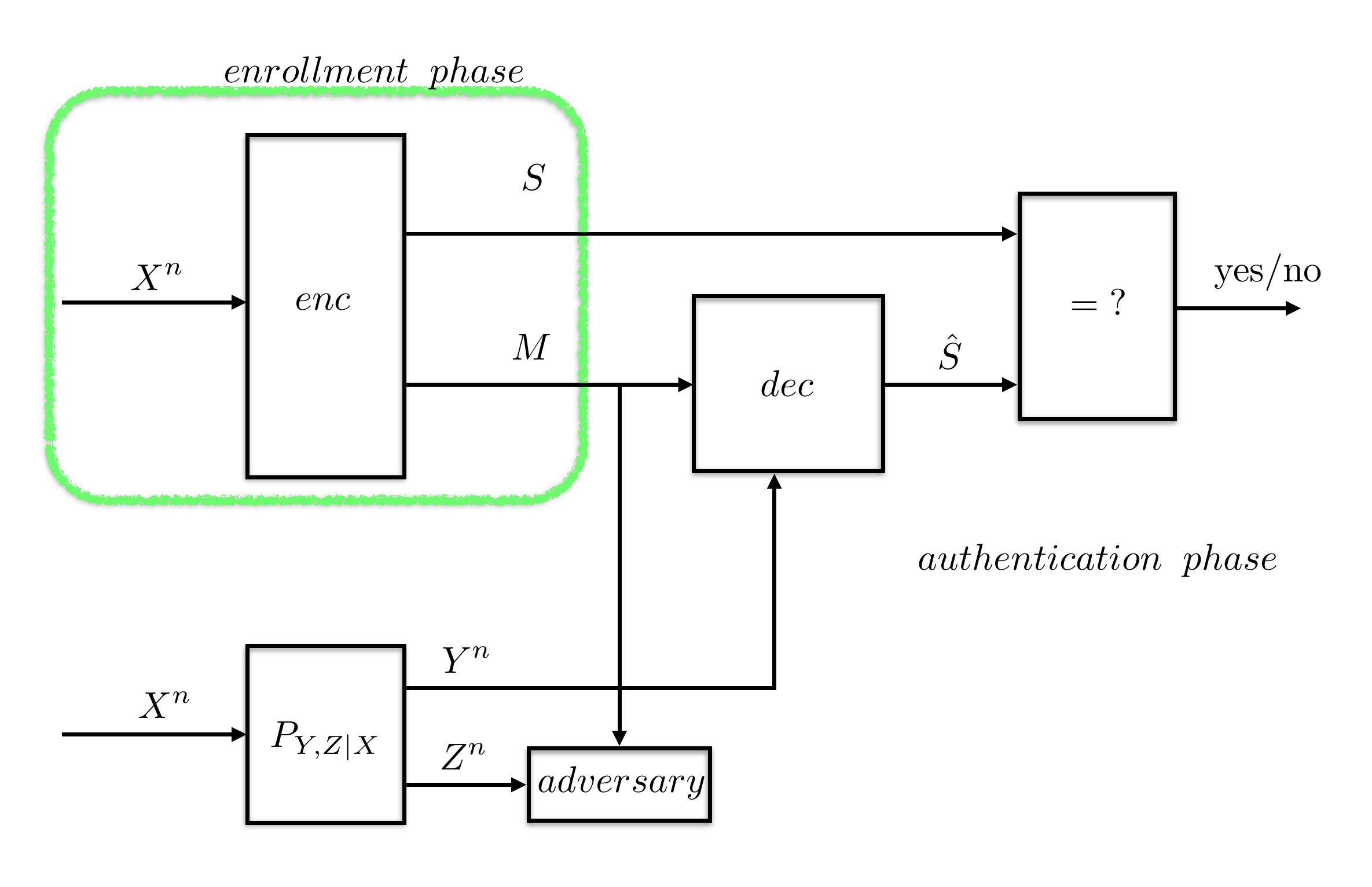}}
\caption{Secret key generation for authentication with a privacy constraint.}
\label{fig:authentication_keyrate}
\end{figure}

ii) When restricting to the case without secret key-based authentication ($R_S=0$), the problem reduces to 
identification with a privacy constraint. By setting $V=U$ in $\mathcal{R}_1$, we obtain the identification-compression-leakage rate 
region consisting of all tuples $(R_I,R_C,L)$ such that
\begin{align*}
R_I &\leq I(Y;U), \\
R_C &\geq R_I +I(X;U|Y),\\
L & \geq I(X;U,Z),
\end{align*}
for some joint distributions of the form $P_{X,Y,Z}P_{U|X}$. Furthermore, without the leakage constraint, this result recovers the optimal 
compression-identification rate (capacity/storage) tradeoff in \cite{Tuncel}, \cite{TuncelGunduz}.

iii) When there is no compression rate constraint (i.e., $R_C = H(X)$) and, furthermore, the adversary has no ``on-line'' side information 
(i.e., $Z = \varnothing$), the region reduces to the set of all tuples $(R_I,L,R_S)$ such that
\begin{align*}
R_I &\leq I(Y;U), \\
L & \geq I(X;V,Y) - I(X;Y|U)= I(X;V|Y)+I(Y;U),\\
R_S &\leq I(V;Y|U)=I(Y;V)-I(Y;U),
\end{align*}
for some joint distributions of the form $P_{X,Y} P_{V|X}P_{U|V}$. 
By setting $R_I = I(Y;U)$ in the expression above (thus restricting the region to a potentially smaller set), 
we recover the result in \cite{IgnatenkoWillems_iden}, i.e., we obtain an achievable region that coincides with the region 
derived in  \cite{IgnatenkoWillems_iden}.
\end{remark}

\begin{IEEEproof}[Proof of Theorem \ref{theorem:region_keyrate}]
Achievability is proved based on a random coding argument where we use the definitions and properties 
of $\epsilon$-typicality as in \cite{ElGamalKim}. 

\textbf{Achievability}: 
Our achievable scheme utilizes  layered coding, binning, and subbinning 
as illustrated in Fig. \ref{fig:layered_binning}.\footnote{Intuition for our achievable scheme is as follows. We use two layers of codewords $\{U^n\}$ and $\{V^n\}$ to be able to adapt to the presence of the 
adversary by controlling the information leakage via the descriptions $M$. 
Since the decoder has side information $Y^n$, binning is used to reduce the compression rate at each layer. This  also essentially reduces the information leakage rate. Moreover, we divide the second layered bin into subbins for the secret key in order to prevent the key leakage. 
We note that the availability of side information $Z^n$ at the adversary has an impact on the structure of the achievable scheme. 
If $Z^n$ becomes degenerate, i.e., $Z=\emptyset$, then it can be shown that the second layer of codewords and 
subbinning are not required to achieve the optimal identification-compression-leakage-keyrate region.} 
Fix $P_{V|X}$ and $P_{U|V}$. Let $\epsilon$ and $\delta_{\epsilon}$ be positive real numbers where $\delta_{\epsilon} \rightarrow 0$ as $\epsilon \rightarrow 0$. Assume that $I(V;Y|U)-I(V;Z|U) >0$. Note also that the joint distribution $P_{X,Y,Z}P_{V|X}P_{U|V}$ implies that $U-V-X-(Y,Z)$ forms a Markov chain.

\begin{figure}[ht]
	\centerline{\includegraphics[width=17cm]{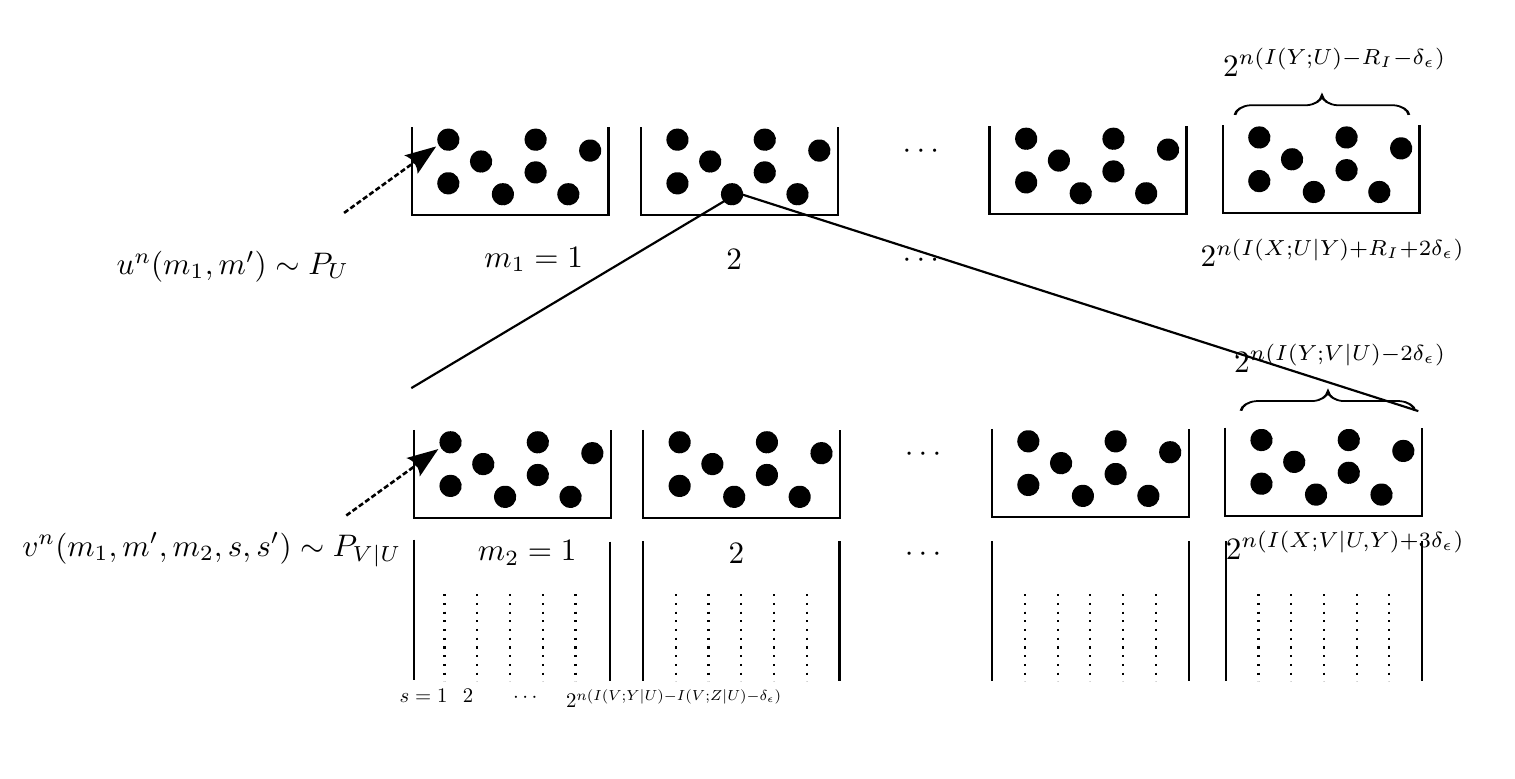}}
	\caption{Layered binning; rate allocation between compression and identification only applies to the first layered codewords, i.e., $m_1 \in [1:2^{n(I(X;U|Y)+R_I+2\delta_{\epsilon})}]$. The bin indices $(m_1,m_2)$ are sent to the decoder as a helping data. The secret key is chosen as a subbin index $s$ of the chosen codeword $v^n$. }
	\label{fig:layered_binning}
\end{figure}

 1) \emph{Codebook generation:} Randomly and independently generate codewords $u^n(j)$ for $j \in [1:2^{n(I(X;U) + \delta_{\epsilon})}]$, 
according to the product distribution $\prod_{i=1}^n P_U(u_i)$. Choosing some identification rate
 \begin{equation}\label{eq:binning_condition}
   R_I \leq I(U;Y) -\delta_{\epsilon}, 
 \end{equation}
we distribute the codewords uniformly at random into $2^{n(I(X;U|Y) +R_I + 2\delta_{\epsilon})}$ bins $b_U(m_1)$, $m_1 \in [1:2^{n(I(X;U|Y) + R_I + 2\delta_{\epsilon})}]$.  Each bin contains $2^{n(I(U;Y) - R_I -\delta_{\epsilon})}$ codewords, each indexed by $m'$, where
$I(X;U) - I(X;U|Y) = I(U;Y)$ follows from the fact that $U-X-Y$.
There exists a one-to-one mapping between index $j$ and the pair of bin/codeword indices $(m_1,m')$ 
such that, without loss of generality, we can identify $j=(m_1,m')$.

For each $j$, randomly and conditionally independently generate codewords 
$v^n(j,k)$ for $k \in [1:2^{n(I(X;V|U) + \delta_{\epsilon})}]$, 
according to the conditional product distribution $\prod_{i=1}^nP_{V|U}(v_i|u_i(j))$, 
and distribute these codewords uniformly at random into $2^{n(I(X;V|U,Y) + 3\delta_{\epsilon})}$ bins $b_V(j,m_2)$, 
$m_2 \in [1:2^{n(I(X;V|U,Y) + 3\delta_{\epsilon})}]$. 
Each bin $b_V(j,m_2)$ contains $2^{n(I(V;Y|U) - 2\delta_\epsilon)}$ codewords, where
$I(X;V|U) - I(X;V|U,Y) = I(V;Y|U)$ follows from the fact that $U-V-X-Y$.
Moreover, the codewords of each bin $b_V(j,m_2)$ are distributed uniformly at random 
into subbins, indexed by $s$, where $s \in [1:2^{n(I(V;Y|U) - I(V;Z|U)-\delta_{\epsilon})}]$. 
The index $s$ here represents a subbin index of the second-layered bin. 
In each subbin, there are $2^{n(I(V;Z|U)-\delta_{\epsilon})}$ codewords, each indexed by $s'$. 
There exists a one-to-one mapping between index $k$ and the triple of bin/subbin/codeword indices $(m_2,s,s')$
such that, without loss of generality, we can identify $k=(m_2,s,s')$.

 2) \emph{Enrollment:} For each user $w\in [1:K]$, given $X^n(w) = x^n(w)$, the encoder looks for $u^n(j)$ that is jointly typical 
 with $x^n(w)$ and then for $v^n(j,k)$ that is jointly typical with $(x^n(w),u^n(j))$. From the covering lemma \cite{ElGamalKim}, with high probability, there exist such codeword pairs since there are more than $2^{nI(X;U)}$ codewords $u^n(j)$ and, for each $j$, 
 there are more than $2^{nI(X;V|U)}$ codewords $v^n(j,k)$. If there are more than one such pairs, the encoder selects one of them uniformly at random. 
 Let the chosen codeword indices of user $w$ be denoted by $(j(w),k(w))=(m_1(w),m'(w),m_2(w),s(w),s'(w))$.
 The encoder stores the corresponding bin indices $m_1(w)$ and $m_2(w)$ into the database as the stored description of user $w$. The compression rate of each user is thus given by
 \begin{align}
   R_C &= I(X;U|Y)+R_I+I(X;V|U,Y) + 5\delta_{\epsilon} \nonumber \\
    &= I(X;V|Y)+R_I + 5\delta_{\epsilon}, \label{eq:rate_condition}
 \end{align}
 where the second equality follows from the chain rule
$I(X;U|Y) +  I(X;V|U,Y) = I(X;U,V|Y)$ and from the fact that $I(X;U,V|Y) = I(X;V|Y)$ due to $U-(V,Y)-X$.
The secret key corresponding to user $w$ is given by the subbin index $s(w)$ in which the chosen sequence 
$v^n(j(w),k(w)) \in  b_V(j(w),m_2(w))$ is found.

3) \emph{Identification/Authentication:} Given an arbitrary user $W \in [1:K]$, let $y^n$ denote the realization of the corresponding measurement sequence
$Y^n$.

The decoder has bin indices $(\bar{m}_1,\bar{m}_2)=((m_1(1),\ldots,m_1(K)),(m_2(1),\ldots,m_2(K)))$ and observes $y^n$. 
Then, for each $w \in [1:K]$, it looks into the corresponding bins $b_U(m_1(w))$ and $b_V(m_1(w),m',m_2(w))$ for all indices
$m'$ forming bin $b_V(m_1(w),m',m_2(w))$, and check if there exists a codeword pair  
\[ (u^n(m_1(w),m'), v^n(m_1(w), m', m_2(w),s,s')) \] 
jointly typical with $y^n$ for some $m', s, s'$. Suppose that there exists a unique $\hat{w}$ for which this condition holds. Then, the decoder outputs the identified user $\hat{w}$. 
Otherwise, if none or more than one user index satisfy the condition, an identification failure is declared. 
Suppose that such unique $\hat{w}$ is found. Then, the decoder outputs also $\hat{s}$ to be the $s$-index of one of the 
codeword pairs 
\[ (u^n(m_1(\hat{w}),m'), v^n(m_1(\hat{w}), m', m_2(\hat{w}),s,s')) \]
satisfying the typicality condition. If there exist more than one
such indices $s$, one is chosen at random. Finally, the decoder compares $\hat{s}$ with $s(\hat{w})$, and declares the identified user as 
successfully authenticated if $\hat{s} = s(\hat{w})$.

Let $U^n(M_1(w),M')$ and $V^n(M_1(w), M', M_2(w),S(w),S')$ be the codewords chosen at the encoder for each $w \in [1:K]$ in the enrollment phase, and $(\bar{M}_1,\bar{M}_2)$ be the corresponding  index vectors of the bins stored in the database.
We note that by symmetry of the codebook generation, the analysis of identification/authentication error does not depend on which user $W \in [1:K]$ is present to the system. Suppose that a user $W=w$ is present. The relevant identification/authentication error events are:
\begin{align*}
E_0 &: \{(Y^n,U^n(M_1(w),M'),V^n(M_1(w), M', M_2(w),S(w),S')) \notin \mathcal{T}_{\epsilon}^{(n)}(Y,U,V) \} \\
E_{\rm Id}&: \{(Y^n,U^n(M_1(\hat{w}),m'),V^n(M_1(\hat{w}), m', M_2(\hat{w}),s,s')) \in \mathcal{T}_{\epsilon}^{(n)}(Y,U,V)\ \text{for some}\ \hat{w} \neq w\ \text{and}\ m',s,s' \} \\
E_{\rm Au}&: \{(Y^n,U^n(M_1(w),M'),V^n(M_1(w), M', M_2(w),\hat{s},s')) \in \mathcal{T}_{\epsilon}^{(n)}(Y,U,V)\ \text{for some}\ \hat{s} \neq S(w)\ \text{and}\  s' \}. 
\end{align*}

For any $W=w$, by LLN, $(X^n,U^n(M_1(w),M'),V^n(M_1(w), M', M_2(w),S(w),S'),Y^n,Z^n)$ are jointly typical with high probability. Thus, $\PP(E_0) \rightarrow 0$ as $n \rightarrow \infty$. From the packing lemma \cite{ElGamalKim}, $\PP(E_{\rm Id} \cup E_{\rm Au} )\rightarrow 0$ as $n \rightarrow \infty$ if $\frac{1}{n}\log K\cdot |\mathcal{M}'||\mathcal{S}||\mathcal{S}'| < I(Y;U,V)$ and $\frac{1}{n}\log|\mathcal{S}||\mathcal{S}'| < I(Y;V|U)$, where $K=2^{n(R_I-\delta)}$. These conditions are satisfied by the code construction.

Note that here we show that the average probability of identification/authentication error $\PP((\hat{W},\hat{S}) \neq (W,S(W))$ can be made arbitrary small for $n$ sufficiently large. Using the expurgation argument with respect to the user index $W$, it can be shown that the maximal error probability of identification/authentication $\max_{W \in \mathcal{W}^{(n)}}\PP((\hat{W},\hat{S}) \neq (W,S(W))$ can also be made arbitrary small at the same asymptotic identification rate $R_I$.

Before proceeding with the analysis of leakage rate, we give a lemma which provides a bound on the $n$-letter conditional entropy based on properties of jointly typical sequences. 
\begin{lemma} \label{lemma:1}
Let $J(w)$ be the index of codeword $U^n$. If $\PP((U^n(J(w)),Z^n) \in \mathcal{T}_{\epsilon}^{(n)}) \rightarrow 1$ as $n \rightarrow \infty$, we have that $H(Z^n|J(w)) \leq n(H(Z|U)+\delta_{\epsilon})$.
\end{lemma}
\begin{IEEEproof}
The proof is given in Appendix \ref{appendix:Lemma1}.
\end{IEEEproof}

\textit{Information leakage analysis}: 
For any $W=w\in[1:K]$, the information leakage averaged over all randomly chosen codebook $\mathcal{C}_n$ can be bounded as follows.
\begin{align*}
&I(X^n(w);\bar{M}_1,\bar{M}_2,Z^n|\mathcal{C}_n) \\
&\overset{(a)}{=} I(X^n(w);M_1(w),M_2(w),Z^n|\mathcal{C}_n)\\
&= H(X^n(w))- H(X^n(w)|M_1(w),M_2(w),Z^n,\mathcal{C}_n)\\
&\leq nH(X)- H(X^n(w)|J(w),Z^n,\mathcal{C}_n)+H(M_2(w)|\mathcal{C}_n)\\
&\leq nH(X)- H(X^n(w),Z^n) + H(J(w)|\mathcal{C}_n) +H(Z^n|J(w),\mathcal{C}_n) +H(M_2(w)|\mathcal{C}_n)\\
&\overset{(b)}{\leq} -nH(Z|X) + n(I(X;U)+\delta_{\epsilon}) + n(H(Z|U)+\delta_{\epsilon}) +n(I(X;V|U,Y)+3\delta_{\epsilon})\\
&\overset{(c)}{\leq} n(I(X;V,Y)-I(X;Y|U)+I(X;Z|U)+ \delta_{\epsilon}'),
\end{align*}
 where $(a)$ follows since given the codebook, $X^n(w)-(M_1(w),M_2(w),Z^n)-(\bar{M}_1^{\setminus w},\bar{M}_2^{\setminus w})$ forms a Markov chain, $(b)$ follows from the memoryless property of the sources, from the codebook generation where $J(w) \in [1:2^{n(I(X;U) + \delta_{\epsilon})}]$ and $M_2(w) \in [1:2^{n(I(X;V|U,Y) + 3\delta_{\epsilon})}]$, and from bounding the term $H(Z^n|J(w))$ as in Lemma \ref{lemma:1}, and $(c)$ follows from the Markov chain $U-V-X-(Y,Z)$ for some $\delta_{\epsilon}' \geq 5\delta_{\epsilon}$. The information leakage constraint  is satisfied if
 \begin{equation}\label{eq:leakage_condition}
   L \geq I(X;V,Y)-I(X;Y|U)+I(X;Z|U).
 \end{equation}

\textit{Key rate analysis}: 
For any $W=w\in[1:K]$, we consider the following bound on the secret key rate.
 \begin{align*}
&H(S(w)|\mathcal{C}_n) \geq H(S(w)|J(w),M_2(w),S'(w),\mathcal{C}_n)\\
&= H(S(w),J(w),M_2(w),S'(w)|\mathcal{C}_n)-H(J(w),M_2(w),S'(w)|\mathcal{C}_n) \\
&\overset{(a)}{\geq} H(U^n,V^n|\mathcal{C}_n)-H(J(w)|\mathcal{C}_n)-H(M_2(w)|\mathcal{C}_n)-H(S'(w)|\mathcal{C}_n)\\
&\overset{(b)}{\geq} n(I(X;U,V)-2\delta_{\epsilon}) - n(I(X;U)+\delta_{\epsilon}) -n(I(X;V|U,Y)+3\delta_{\epsilon})-n(I(V;Z|U)-\delta_{\epsilon})\\
& \geq n(I(Y;V|U)-I(Z;V|U) -\delta_{\epsilon}'),
\end{align*}
 where $(a)$ follows since given the codebook, codewords  $(U^n,V^n)$ are functions of $(J(w),K(w))=(J(w),M_2(w),S(w),S'(w))$  and $(b)$ follows from the codebook generation where $J(w) \in [1:2^{n(I(X;U) + \delta_{\epsilon})}]$ and $M_2(w) \in [1:2^{n(I(X;V|U,Y) + 3\delta_{\epsilon})}]$ and since probability of a specific pair $(u^n,v^n)$ being selected in the enrollment can be bounded by $p(u^n,v^n) \leq \sum_{x^n(w) \in \mathcal{T}_{\epsilon}^{(n)}(X|u^n,v^n)}p(x^n(w))\leq 2^{-n(I(X;U,V)-2\delta_{\epsilon})}$, where the last inequality follows from properties of jointly typical sequences. 
 Therefore, the key rate constraint  is satisfied if
  \begin{equation}\label{eq:keyrate_condition}
   R_S \leq I(Y;V|U)-I(Z;V|U).
 \end{equation}

\textit{Key leakage analysis}: 
For any $W=w\in[1:K]$,  the key leakage averaged over all possible codebooks can be bounded as follows.
 \begin{align}
&I(S(w);\bar{M}_1,\bar{M}_2,Z^n|\mathcal{C}_n) \overset{(a)}{=}  I(S(w);M_1(w),M_2(w),Z^n|\mathcal{C}_n)\nonumber\\
&\leq H(S(w)|\mathcal{C}_n)-H(S(w)|J(w),M_2(w),Z^n,\mathcal{C}_n)\nonumber\\
&= H(S(w)|\mathcal{C}_n)-H(S(w),J(w),M_2(w),Z^n|\mathcal{C}_n)+H(J(w),M_2(w),Z^n|\mathcal{C}_n)\nonumber\\
&\leq H(S(w)|\mathcal{C}_n) -H(S(w),J(w),M_2(w),Z^n,S'(w)|\mathcal{C}_n) +H(S'(w)|S(w),J(w),M_2(w),Z^n,\mathcal{C}_n)\nonumber\\
&\qquad+H(J(w)|\mathcal{C}_n)+H(M_2(w)|\mathcal{C}_n)+H(Z^n|J(w),\mathcal{C}_n)\nonumber\\
&\overset{(b)}{\leq} H(S(w)|\mathcal{C}_n)-H(U^n,V^n,Z^n|\mathcal{C}_n) + n\epsilon_n +H(J(w)|\mathcal{C}_n)+H(M_2(w)|\mathcal{C}_n)+H(Z^n|J(w),\mathcal{C}_n)\nonumber\\
&\overset{(c)}{\leq} H(S(w)|\mathcal{C}_n) -n(I(X;U,V)+H(Z|U,V)-2\delta_{\epsilon}) + n\epsilon_n + n(I(X;U)+\delta_{\epsilon}) \nonumber\\ &\qquad +n(I(X;V|U,Y)+3\delta_{\epsilon})+n(H(Z|U)+\delta_{\epsilon}) \overset{(d)}{\leq} n\delta_{\epsilon}'', \label{eq:keyleakage_condition}
\end{align}
where $(a)$ follows since given the codebook, $S(w)-(M_1(w),M_2(w),Z^n)-(\bar{M}_1^{\setminus w},\bar{M}_2^{\setminus w})$ forms a Markov chain, $(b)$ follows since given the codebook, codewords $(U^n,V^n)$ are functions of $(J(w),K(w))=(J(w),M_2(w),S(w),S'(w))$, and from the Fano's inequality $H(S'(w)|S(w),J(w),M_2(w),Z^n) \leq n\epsilon_n$ which holds because from the codebook generation, the number of possible codewords $V^n$ for a given $(J(w),M_2(w),S(w))$ is less than $2^{nI(V;Z|U)}$ and therefore with high probability $V^n$ (and thus $S'(w)$) can be decoded given $(S(w),J(w),M_2(w),U^n,Z^n)$, $(c)$ follows from bounding the term $H(U^n,V^n,Z^n)$ using properties of jointly typical sequences, i.e., 
\begin{align*}
p(u^n,v^n,z^n) &\leq \sum_{x^n(w) \in \mathcal{T}_{\epsilon}^{(n)}(X|u^n,v^n,z^n)}p(x^n(w),z^n)\\& \leq 2^{-n(H(X,Z)-H(X|U,V,Z)-2\delta_{\epsilon})}=2^{-n(I(X;U,V)+H(Z|U,V)-2\delta_{\epsilon})}, 
\end{align*} 
where the equality holds since $Z-X-(U,V)$, 
from the codebook generation, and from Lemma~\ref{lemma:1}, and finally  $(d)$ follows from the codebook generation where  $S(w) \in [1:2^{n(I(Y;V|U)-I(Z;V|U)-\delta_{\epsilon})}]$.

 Given \eqref{eq:keyleakage_condition}, combining \eqref{eq:binning_condition} to \eqref{eq:keyrate_condition} and invoking the random coding argument complete the achievability proof.

\textbf{Converse}: We prove the converse for the average probability of error with respect to a user $W$ randomly selected with uniform probability over the set $[1:K]$. Clearly, the average probability of error is less than or equal to the maximal probability of error over the users, such that the achievable region with respect to this less restrictive criterion contains the one with respect to the criterion given in our definition. The converse here implies that the two regions match and therefore completes the proof of the theorem. 

Conditioned on $W=w$, the joint PMF of all relevant random variables is given by
\begin{align*}
&P_{X^n(1),\dots,X^n(K),M(1),\ldots,M(K),S(1),\ldots,S(K),Y^n,Z^n|W=w}\\
&=P_{X^n(w),M(w),S(w),Y^n,Z^n|W=w} \prod_{j=1, j \neq w}^K P_{X^n(j),M(j),S(j)|W=w}\\
&=P_{X^n(w),M(w),S(w)|W=w}P_{Y^n,Z^n|X^n(w)} \prod_{j=1, j \neq w}^K P_{X^n(j),M(j),S(j)|W=w},
\end{align*}
where $P_{X^n(j)}(x^n) = \prod_{i=1}^n P_{X}(x_i)$ for $j \in [1:K]$, and $P_{Y^n,Z^n|X^n(w)}(y^n,z^n|x^n) = \prod_{i=1}^n P_{Y,Z|X}(y_i,z_i|x_i)$. 

Let us define $U_i \triangleq (W,M(W),Y_{i+1}^n,Z^{i-1})$ and $V_i \triangleq (W,M(W),S(W),Y_{i+1}^n,Z^{i-1})$ which satisfy $U_i-V_i-X_i(W)-(Y_i,Z_i)$ for all $i=1,\ldots,n$. This can be seen as $U_i$ is included in $V_i$ and $(Y_i,Z_i)$ is independent of $V_i$ given $X_i(W)$ due to the memoryless property of the ``channel'' $P_{Y,Z|X}$.  For any achievable tuple $(R_I,R_C,L,R_S) \in \mathbb{R}^4_{+}$, we have Fano's inequality $H(W,S(W)|\bar{M},Y^n) \leq n\epsilon_n$.

It then follows that
\begin{align}
n(R_I - \delta_n) &\leq H(W)= H(W|\bar{M},Y^n) + I(W;\bar{M},Y^n)\nonumber\\
&\overset{(a)}{\leq}  n\epsilon_n+ I(W;\bar{M},Y^n)\nonumber\\
&\overset{(b)}{=} n\epsilon_n+ I(W;Y^n|\bar{M})\nonumber\\
&\leq n\epsilon_n+ H(Y^n)-H(Y^n|W,\bar{M})\nonumber\\
&\overset{(c)}{=}  n\epsilon_n+ H(Y^n)-H(Y^n|W,M(W))\label{eq:R_I}\\
&\leq \sum_{i=1}^n H(Y_i)-H(Y_i|W,M(W),Y_{i+1}^n,Z^{i-1}) + n\epsilon_n,\nonumber\\
&\overset{(d)}{=} \sum_{i=1}^n I(Y_i;U_i) + n\epsilon_n,\nonumber
\end{align}
where $(a)$ follows from Fano's inequality $H(W|\bar{M},Y^n) \leq H(W,S(W)|\bar{M},Y^n) \leq n\epsilon_n$, $(b)$ follows from the fact that $W$ is independent of $\bar{M}$, $(c)$ follows from the fact that conditioned on $W=w$, we have that $Y^n-M(w)-\bar{M}^{\setminus w}$ forms a Markov chain (see Appendix \ref{appendix:markovchain} (I) for the proof), and $(d)$ follows from the definition of $U_i$. 

Next,  
\begin{align*}
n(R_C-R_I + \delta_n) &\geq H(M(W))-H(W) \\
&\overset{(a)}{\geq} H(M(W))-I(Y^n;W,M(W))-n\epsilon_n\\
&= -H(W|M(W))+ H(W,M(W)|Y^n)-n\epsilon_n\\
&\geq -H(W|M(W))+ H(W,M(W),S(W)|\bar{M}^{\setminus W},Y^n)
-H(S(W)|\bar{M},W,Y^n)\\&\qquad-H(M(W),S(W)|\bar{M}^{\setminus W},Y^n,X^n(W),Z^n)-n\epsilon_n\\
&\overset{(b)}{=} -H(W|\bar{M},X^n(W),S(W))+ P\\
&\overset{(c)}{=} -H(W|\bar{M},X^n(W),S(W),Y^n,Z^n)+ P
\end{align*}
where $(a)$ follows from \eqref{eq:R_I}, $(b)$ follows since $W$ is independent of $(\bar{M},X^n(W),S(W))$ and the definition $P \triangleq H(W,M(W),S(W)|\bar{M}^{\setminus W},Y^n)-H(M(W),S(W)|\bar{M}^{\setminus W},Y^n,X^n(W),Z^n)-n\epsilon_n-H(S(W)|\bar{M},W,Y^n)$, and $(c)$ follows from the fact that we have $I(W;Y^n,Z^n|\bar{M},X^n(W),S(W))=0$ or equivalently $H(Y^n,Z^n|\bar{M},X^n(W),S(W))-H(Y^n,Z^n|\bar{M},X^n(W),S(W),W) \leq 0$ which holds since i) conditioned on $W=w$, $(Y^n,Z^n)-X^n(w)-(\bar{M},S(w))$ forms a Markov chain (see Appendix \ref{appendix:markovchain} (II)) and ii) we have the Markov chain $W-X^n(W)-(Y^n,Z^n)$ derived from the given ``channel'' $P_{Y,Z|X}$.

Continuing the chain of inequalities and substituting the value of $P$, we get
\begin{align*}
n(R_C-R_I + \delta_n) &\geq  H(W,M(W),S(W)|\bar{M}^{\setminus W},Y^n)
-H(S(W)|\bar{M},W,Y^n)\\ &\qquad -H(W,M(W),S(W)|\bar{M}^{\setminus W},Y^n,X^n(W),Z^n)-n\epsilon_n\\
&\overset{(d)}{\geq} I(W,M(W),S(W);X^n(W),Z^n|\bar{M}^{\setminus W},Y^n)-2n\epsilon_n\\
&\overset{(e)}{\geq} H(X^n(W),Z^n|W,Y^n)-H(X^n(W),Z^n|W,\bar{M},S(W),Y^n)-2n\epsilon_n\\
&\overset{(f)}{=} H(X^n(W),Z^n|Y^n)-H(X^n(W),Z^n|W,\bar{M},S(W),Y^n)-2n\epsilon_n\\
&\geq \sum_{i=1}^n H(X_i(W),Z_i|Y_i) -H(X_i(W),Z_i|W,M(W),S(W),Y_{i+1}^n,Z^{i-1},Y_i)-2n\epsilon_n\\
&\overset{(g)}{\geq}  \sum_{i=1}^n I(X_i(W);V_i|Y_i)-2n\epsilon_n,
\end{align*}
where  $(d)$ follows from Fano's inequality where $H(S(W)|\bar{M},W,Y^n)\leq H(W,S(W)|\bar{M},Y^n)\leq n\epsilon_n$, $(e)$ follows from the fact that $H(X^n(W),Z^n|\bar{M}^{\setminus W},Y^n) \geq H(X^n(W),Z^n|\bar{M}^{\setminus W},Y^n,W)$ and that conditioned on $W=w$, we have the Markov chain $(X^n(w),Z^n)-Y^n-\bar{M}^{\setminus w}$ (see Appendix \ref{appendix:markovchain} (III)), $(f)$ follows from the facts that $W$ is independent of $X^n(W)$ and the Markov chain $W-X^n(W)-(Y^n,Z^n)$, and finally $(g)$ follows from the definition of $V_i$.

The information leakage can be bounded as follows.
\begin{align*}
&n(L + \delta_n)\geq \max_{W \in \mathcal{W}^{(n)}} I(X^n(W);\bar{M},Z^n) 
\geq \frac{1}{K}\sum_{w=1}^K I(X^n(w);\bar{M},Z^n|W=w)\\
&= I(X^n(W);\bar{M},Z^n|W)\\
&\overset{(a)}{=}  I(X^n(W);W,\bar{M},Z^n)\\
&= I(X^n(W);W,\bar{M},S(W),Y^n)-I(X^n(W);S(W)|W,\bar{M},Y^n)\\&\qquad -I(X^n(W);Y^n|W,\bar{M})+I(X^n(W);Z^n|W,\bar{M})\\
&\overset{(b)}{\geq} I(X^n(W);W,\bar{M},S(W),Y^n)-n\epsilon_n -I(X^n(W);Y^n|W,\bar{M})+I(X^n(W);Z^n|W,\bar{M})\\
&\overset{(c)}{\geq} I(X^n(W);W,M(W),S(W),Y^n)-n\epsilon_n -I(X^n(W);Y^n|W,M(W))+I(X^n(W);Z^n|W,M(W)),
\end{align*}
where $(a)$ follows from the fact that $X^n(W)$ is independent of $W$, $(b)$ follows from Fano's inequality, $H(S(W)|W,\bar{M},Y^n)\leq H(W,S(W)|\bar{M},Y^n) \leq n\epsilon_n$, and $(c)$ follows from the fact that conditioned on $W=w$, we have the Markov chain $(X^n(w),Y^n,Z^n,S(w))-M(w)-\bar{M}^{\setminus w}$ (see Appendix \ref{appendix:markovchain} (IV)).

Continuing the chain of inequalities, we have
\begin{align*}
&n(L + \delta_n) \geq \sum_{i=1}^n H(X_i(W))-H(X_i(W)|W,M(W),S(W),X^{i-1}(W),Y^n)  - H(Y_i|W,M(W),Y_{i+1}^n)\\ &\qquad +H(Y_i|W,M(W),Y_{i+1}^n,X^n(W))+H(Z_i|W,M(W),Z^{i-1}) - H(Z_i|W,M(W),Z^{i-1},X^n(W)) -n\epsilon_n  \\
&\overset{(d)}{\geq} \sum_{i=1}^n H(X_i(W)) -H(X_i(W)|W,M(W),S(W),X^{i-1}(W),Y^n,Z^{i-1}) -I(Y_i;X_i(W))\\ &\qquad + I(Y_i;W,M(W),Y_{i+1}^n) +I(Z_i;X_i(W))-I(Z_i;W,M(W),Z^{i-1}) -n\epsilon_n  \\
&\overset{(e)}{\geq} \sum_{i=1}^n I(X_i(W);W,M(W),S(W),Y_i^n,Z^{i-1}) -I(Y_i;X_i(W)) +I(Z_i;X_i(W))  \\ &\qquad + I(Y_i;W,M(W),Z^{i-1},Y_{i+1}^n)-I(Z_i;W,M(W),Z^{i-1},Y_{i+1}^n)-n\epsilon_n \\
&\overset{(f)}{=} \sum_{i=1}^n I(X_i(W);V_i,Y_i) -I(Y_i;X_i(W)|U_i) +I(Z_i;X_i(W)|U_i) -n\epsilon_n,
\end{align*}
where  $(d)$ follows from the fact that conditioned on $W=w$, we have the Markov chains $X_i(w)-(M(w),S(w),X^{i-1}(w),Y^n)-Z^{i-1}$ and $(Y_i,Z_i)-X_i(w)-(M(w),Y_{i+1}^n,Z^{i-1},X^{n\setminus i}(w))$ which hold due to the memoryless properties of the ``channel'' $P_{Y^n,Z^n|X^n(w)}(y^n,z^n|x^n)=\prod_{i=1}^n P_{Y,Z|X}(y_i,z_i|x_i)$ and from the Markov chain $(Y_i,Z_i)-X_i(W)-W$,
$(e)$ follows from the Csisz\'{a}r's sum identity \cite{CsiszarBook} which in this case is $\sum_{i=1}^n I(Y_i;Z^{i-1}|W,M(W),Y_{i+1}^n) - I(Z_i;Y_{i+1}^n|W,M(W),Z^{i-1}) =0$, and finally
$(f)$ follows from the definitions of $U_i$ and $V_i$, and the Markov chain $U_i-X_i(W)-(Y_i,Z_i)$.

Lastly, the secret key rate can be bounded as follows.
\begin{align}
&n(R_s-\delta_n)  \leq \min_{W \in \mathcal{W}^{(n)}} H(S(W)) 
\leq \frac{1}{K}\sum_{w=1}^K H(S(w)|W=w)= H(S(W)|W)\nonumber\\
&=H(S(W)|W,\bar{M},Z^n)+I(S(W);\bar{M},Z^n|W)\nonumber\\
&=H(S(W)|W,\bar{M},Z^n)+ \frac{1}{K} \sum_{w=1}^K  I(S(w);\bar{M},Z^n|W=w)\nonumber\\
&\leq H(S(W)|W,\bar{M},Z^n)+ \max_{W \in \mathcal{W}^{(n)}} I(S(W);\bar{M},Z^n)\nonumber\\
&\overset{(a)}{\leq} H(S(W)|W,\bar{M},Z^n)+n\delta_n \label{eq:Econverse_start} \\
&\overset{(b)}{\leq} H(S(W)|W,\bar{M},Z^n)-H(S(W)|W,\bar{M},Y^n)+n\delta_n +n\epsilon_n \nonumber\\
&= I(S(W);Y^n|W,\bar{M})- I(S(W);Z^n|W,\bar{M})+n\delta_n +n\epsilon_n \nonumber \\
&\overset{(c)}{=} I(S(W);Y^n|W,M(W))- I(S(W);Z^n|W,M(W)) +n\delta_n +n\epsilon_n \nonumber\\
&= \sum_{i=1}^n I(S(W);Y_i|W,M(W),Y_{i+1}^n) -I(S(W);Z_i|W,M(W),Z^{i-1})+n\delta_n +n\epsilon_n \nonumber\\
&\overset{(d)}{=} \sum_{i=1}^n I(S(W);Y_i|W,M(W),Y_{i+1}^n,Z^{i-1})-I(S(W);Z_i|W,M(W),Y_{i+1}^n,Z^{i-1}) +n\delta_n +n\epsilon_n \nonumber\\
&\overset{(e)}{=} \sum_{i=1}^n I(V_i;Y_i|U_i)-I(V_i;Z_i|U_i)+n\delta_n +n\epsilon_n,\label{eq:Econverse_end}
\end{align}
where $(a)$ follows from the key leakage constraint, $(b)$ follows from Fano's inequality, $(c)$ follows from the fact that conditioned on $W=w$, we have the Markov chain $(S(w),Y^n,Z^n)-M(w)-\bar{M}^{\setminus w}$ (cf. Appendix \ref{appendix:markovchain} (IV)), $(d)$ follows from the Csisz\'{a}r's sum identity,
and $(e)$ follows from the definitions of $U_i$ and $V_i$. 

The proof ends with the standard steps for single letterization using a time-sharing random variable and letting $\delta_n, \epsilon_n \rightarrow 0$ as $n\rightarrow \infty$. The cardinality bounds on the sets $\mathcal{U}$ and $\mathcal{V}$ 
can be proved using the support lemma \cite{CsiszarBook}, and is shown in Appendix \ref{appendix:cardinality}.
\end{IEEEproof}

\subsection{Binary Example}
To demonstrate the derived tradeoff in Theorem \ref{theorem:region_keyrate}, we consider simple binary examples of the special cases in Remark \ref{remark:special_cases} i) and ii) where the Markov chain $X-Y-Z$ holds, i.e., $X\sim \text{Bernoulli}(1/2)$, $Y$ is an erased version of $X$ with erasure probability $p$, and $Z$ is an erased version of $Y$ with erasure probability $q$.
\begin{itemize}
	\item[1)] When there is no identification rate constraint, the region $\mathcal{R}_{i,X-Y-Z}$ in Remark \ref{remark:special_cases} i) reduces to the set of all $(R_C,L,R_S)$ such that
	\begin{align*}
	R_C &\geq p(1-h(\alpha)), \\
	L & \geq (1-q)(1-p) + p(1-h(\alpha)),\\
	R_S &\leq q(1-p)(1-h(\alpha)),
	\end{align*}
	for some $\alpha \in [0,1/2]$, where $h(\cdot)$ is the binary entropy function. The proof is given in Appendix \ref{appendix:proof_example} where setting $U=\emptyset$ in Remark \ref{remark:special_cases} i) is optimal. We can see for example the tradeoff between the secret key rate and the leakage rate, i.e., to achieve a high secret key rate, we need to operate at a higher compression rate and also allow higher amount of information leakage.
	
	\item[2)] When there is no key rate constraint, the region $\mathcal{R}_{ii,X-Y-Z}$ in Remark \ref{remark:special_cases} ii) reduces to the set of all $(R_I,R_C,L)$ such that
	\begin{align*}
	R_I &\leq (1-p)(1-h(\alpha)),\\
	R_C &\geq R_I+ p(1-h(\alpha)), \\
	L & \geq 1-h(\alpha)((1-p)q+p),
	\end{align*}
	for some $\alpha \in [0,1/2]$. The proof follows similarly as that of $\mathcal{R}_{i,X-Y-Z}$ and is therefore omitted. We can see a similar tradeoff between the identification rate and the leakage rate, e.g., to achieve  a high identification rate, we pay the cost of having  high information leakage rate.
\end{itemize}

\section{Secure Identification/Authentication with a Privacy Constraint}\label{sectionIII}
In this section we consider a new problem where the adversary is assumed to be active and tries to deceive the identification/authentication system using its own data. The main difference from the previous problem is that we impose a constraint on the false acceptance probability, replacing constraints on the secret key rate and key leakage. 
\subsection{Problem Formulation}\label{sec:problem_setting_mFAP}
Let us now consider a secure identification/authentication system as shown in Fig. \ref{fig:scheme} with an active adversary. Source, measurement, and side information alphabets, $\mathcal{X}, \mathcal{Y}, \mathcal{Z}$ are assumed to be finite. The users' source sequences $X^{n}(w)$ for $w \in \Wc^{(n)} \triangleq [1:K]$ are independent across the users and have i.i.d. components distributed according to  some fixed source distribution $P_X$. 
Measurement sequence and side information $(Y^n,Z^n)$ are assumed to be outputs of the memoryless channel with given transition probability $P_{Y,Z|X}$ and input $X^n(W)$, where $W$ is the index representing an arbitrary unknown user who presents itself to the system for identification/authentication.

The enrollment and identification/authentication phases follow similarly as in Section \ref{sec:problem_setting_keyrate}. 
In the event of an attack, the adversary presents to the decoder its own sequence $\tilde{y}^n \in \mathcal{Y}^n$ generated as a function of $\bar{M}$ and $Z^n$, in order to gain access to the system. In this case, the adversary will first be identified as one of the users according to the decoding function $g_{\text{Id}}^{(n)}(\bar{M},\tilde{y}^n)$. Its estimate of the key is equal to $g_{\text{Au}}^{(n)}(\bar{M},\tilde{y}^n)$ which will then be compared with the original key of the user whom it is identified to be, e.g., $S(g_{\text{Id}}^{(n)}(\bar{M},\tilde{y}^n))$. We define a false acceptance event to be an event that $g_{\text{Au}}^{(n)}(\bar{M},\tilde{y}^n)=S(g_{\text{Id}}^{(n)}(\bar{M},\tilde{y}^n))$. Operationally, it means that the adversary gains access as if it were user $g_{\text{Id}}^{(n)}(\bar{M},\tilde{y}^n)$. 
The maximum false acceptance probability (mFAP) is defined as $\max_{\tilde{y}^n(\bar{M}, Z^n)\in \mathcal{Y}^n}\PP(g_{\text{Au}}^{(n)}(\bar{M},\tilde{y}^n)=S(g_{\text{Id}}^{(n)}(\bar{M},\tilde{y}^n)))$.\footnote{We note that the maximization here is over the functions $\tilde{y}^n(\cdot)$, not over the sequences in $\mathcal{Y}^n$.}
Since the adversary will be identified as one of the users in the database, we are concerned about whether it will also be positively authenticated and therefore wish to minimize the maximum false acceptance probability \textit{exponentially}.

As before, information leakage rate of user $W$ at the adversary who has access to the stored database $\bar{M}$ and
side information $Z^n$ is given by the mutual information rate $I(X^n(W);\bar{M},Z^n)/n$.

We are interested in characterizing the optimal tradeoff between the identification rate, compression rate, information leakage rate, and mFAP exponent.

\begin{definition}\label{def:code_mFAP}
	An $(|\mathcal{M}^{(n)}|,|\mathcal{W}^{(n)}|,n)$-code for secure identification and authentication with a privacy constraint consists of
	\begin{itemize}
		\item A set of stochastic encoders $F_{w}^{(n)}: w \in \mathcal{W}^{(n)}$ such that the $w$-th encoder takes $X^n(w)$ as input and generates $(M(w),S(w)) \in \mathcal{M}^{(n)} \times \mathcal{S}^{(n)}$ according to a conditional PMF $p(m(w),s(w)|x^n(w))$.
  \item A decoder $g_{\rm Id}^{(n)}: {\mathcal{M}^{(n)}}^K \times \mathcal{Y}^{n} \rightarrow \mathcal{W}^{(n)}$, 
  such that the identified user is $\hat{W} =  g_{\rm Id}^{(n)}(\bar{M}, Y^n)$. 
  \item A decoder $g_{\rm Au}^{(n)}: {\mathcal{M}^{(n)}}^K \times \mathcal{Y}^{n} \rightarrow \mathcal{S}^{(n)}$, 
  such that the estimated secret key is  $\hat{S} = g_{\rm Au}^{(n)}(\bar{M}, Y^n)$.  
\end{itemize}
\hfill $\lozenge$
\end{definition}

\begin{definition}  An identification-compression-leakage-mFAP exponent tuple $(R_I,R_C,L,E) \in \mathbb{R}^{4}_{+}$ is said to be \emph{achievable} if, for any $\delta>0$, there exists a sequence of $(|\mathcal{M}^{(n)}|,|\mathcal{W}^{(n)}|,n)$-codes such that  for all sufficiently large $n$, 
	\begin{align}
	\max_{W \in \mathcal{W}^{(n)}}\PP((\hat{W},\hat{S})\neq (W,S(W)))&\leq \delta, \label{eq:error_constraint2}\\
	\frac{1}{n}\log\big|\mathcal{W}^{(n)}\big| &\geq R_I-\delta,\label{eq:identification_rate_constraint2}\\
	\frac{1}{n}\log\big|\mathcal{M}^{(n)}\big| &\leq R_C+\delta,\label{eq:compression_rate_constraint2}\\
	\max_{W \in \mathcal{W}^{(n)}}\frac{1}{n}I(X^{n}(W);\bar{M},Z^n) &\leq L+\delta, \label{eq:leakage_constraint2}\\
	\max_{\tilde{y}^n(\bar{M},Z^n)\in \mathcal{Y}^n} \PP(g_{\text{Au}}^{(n)}(\bar{M},\tilde{y}^n)=S(g_{\text{Id}}^{(n)}&(\bar{M},\tilde{y}^n))) \leq 2^{-n(E-\delta)}.\label{eq:mFAP_constraint2}
	\end{align}
	The \emph{identification-compression-leakage-mFAP exponent} region $\mathcal{R}_2$ is defined as the closure of all achievable tuples.
	\hfill $\lozenge$
\end{definition}

\subsection{Result}

\begin{theorem}\label{theorem:region_exponent}
	The  region $\mathcal{R}_2$ for the secure identification/authentication problem defined above is given by a set of all tuples $(R_I,R_C,L,E)\in \mathbb{R}^{4}_{+}$ such that
	\begin{align}
	R_I &\leq I(Y;U), \\
	R_C &\geq R_I +I(X;V|Y),\\
	L & \geq I(X;V,Y) - I(X;Y|U)+I(X;Z|U),\\
	E &\leq I(V;Y|U)-I(V;Z|U),
	\end{align}
	for some  $P_{X,Y,Z}P_{V|X}P_{U|V}$
	with  $|\mathcal{U}| \leq |\mathcal{X}| +4, |\mathcal{V}| \leq (|\mathcal{X}|+4)(|\mathcal{X}|+2)$.
	\hfill $\square$
\end{theorem}

\begin{remark}
	The regions $\mathcal{R}_1$ and $\mathcal{R}_2$ specified in Theorems \ref{theorem:region_keyrate} and \ref{theorem:region_exponent} have the same form. In particular, the maximum mFAP exponent in Theorem \ref{theorem:region_exponent} is equivalent to the maximum achievable secret key rate in Theorem \ref{theorem:region_keyrate}. This reveals a connection between the achievable secret key rate and the security of identification/authentication system in terms of false acceptance probability.
	
	Intuitively, the equivalence follows from the fact that the coding scheme used to prove Theorem \ref{theorem:region_exponent} also achieves negligible key leakage rate for each user, implying that the adversary has no useful knowledge about the secret  key. It can then only guess the secret  key $S$ from possible values in a set whose cardinality is at least $2^{H(S)}$. Therefore, the false acceptance probability is upper-bounded by $2^{-H(S)}$ which is further bounded by $2^{-n(R_s-\delta)}$ when translating to the problem with the secret key rate constraint.
	The same observation holds true when specializing to the single user case \cite{Kittichokechai_b}. This is also noted in  \cite{WillemsIgnatenko} for the case without adversary's side information.
\end{remark}

\begin{IEEEproof}[Proof of Theorem \ref{theorem:region_exponent}] The proof of identification rate, compression rate, and information leakage rate remain the same as in the previous problem in Section \ref{sectionII}. We will only provide the proof of the mFAP exponent of which the main idea follows similarly as that in \cite{WillemsIgnatenko},\cite{Kittichokechai_b}. \par
	\textbf{Achievability}:
	We use the same achievable scheme as in the proof of Theorem \ref{theorem:region_keyrate}.  
	For an achievable mFAP exponent, we consider the adversary who
	knows $\bar{m}=(\bar{m}_1,\bar{m}_2)$ and side information $z^n$. Let $g_{\text{Id}}(\cdot)$ and $g_{\text{Au}}(\cdot)$ denote the decoding functions for identification and the secret key estimation in the achievable scheme. The adversary tries to select a sequence $\tilde{y}^n(\bar{m},z^n)$ that results in the estimated key $g_{\text{Au}}(\bar{m},\tilde{y}^n)$ equal to the original key of the user it is identified to be, i.e., $S(g_{\text{Id}}(\bar{m},\tilde{y}^n))$.
	
	In the achievable scheme, the secret key is chosen as the subbin index of the selected codeword $v^n$. Thus, the adversary only needs to consider the secret key that results from codewords $v^n$ which are jointly typical with $x^n$. There are in total $2^{n(I(X;U,V)+2\delta_{\epsilon})}$ such codewords generated.
	
	From the binning scheme with uniform bin and subbin index assignment, we have that the joint probability that a description $m$ of certain user $\tilde{w}\in\{1,\ldots,K\}$ is selected and a certain secret key of that user $s(\tilde{w})$ is chosen is equal to a total number of jointly typical codewords $v^n$ with corresponding indices $m(\tilde{w})=(m_1(\tilde{w}),m_2(\tilde{w}))$ and $s(\tilde{w})$ divided by a total number of jointly typical codewords $v^n$. That is,
	\begin{align}
	\PP(M(\tilde{w})=m,S(\tilde{w})=s)
	&\leq \frac{\Big\lceil\frac{ \PP(M(\tilde{w})=m)\cdot 2^{n(I(X;U,V)+2\delta_{\epsilon})}}{|\mathcal{S}|}\Big\rceil}{2^{n(I(X;U,V)+2\delta_{\epsilon})}}. \label{eq:joint_prob}
	\end{align}
	Then
	\begin{align*}
	&\mbox{mFAP} = \max_{\tilde{y}^n(\bar{M},Z^n)\in \mathcal{Y}^n} \PP(g_{\text{Au}}(\bar{M},\tilde{y}^n(\bar{M},Z^n)) =S(g_{\text{Id}}(\bar{M},\tilde{y}^n(\bar{M},Z^n))))\\
	&= \max_{\tilde{y}^n(\cdot)\in \mathcal{Y}^n} \sum_{\bar{m}}\sum_{z^n}  \PP(\bar{M}=\bar{m}, Z^n=z^n,g_{\text{Au}}(\bar{m},\tilde{y}^n(\bar{m},z^n)) =S(g_{\text{Id}}(\bar{m},\tilde{y}^n(\bar{m},z^n)))) \\
	&=  \max_{\tilde{y}^n(\cdot)\in \mathcal{Y}^n} \sum_{\bar{m}}\sum_{z^n} \PP(\bar{M}=\bar{m}, S(g_{\text{Id}}(\bar{m},\tilde{y}^n))= g_{\text{Au}}(\bar{m},\tilde{y}^n) )\cdot \PP(Z^n=z^n|\bar{M}=\bar{m}, S(g_{\text{Id}}(\bar{m},\tilde{y}^n))= g_{\text{Au}}(\bar{m},\tilde{y}^n)) \\
	&\overset{(a)}{=} \max_{\tilde{y}^n(\cdot)\in \mathcal{Y}^n} \sum_{\bar{m}}\sum_{z^n} \Big( \PP(\bar{M}^{\setminus g_{\text{Id}}(\bar{m},\tilde{y}^n)}=\bar{m}^{\setminus g_{\text{Id}}(\bar{m},\tilde{y}^n)})\cdot  \PP(M(g_{\text{Id}}(\bar{m},\tilde{y}^n))=m(g_{\text{Id}}(\bar{m},\tilde{y}^n)), S(g_{\text{Id}}(\bar{m},\tilde{y}^n))  = g_{\text{Au}}(\bar{m},\tilde{y}^n) )\cdot \\ &\qquad\PP(Z^n=z^n|\bar{M}=\bar{m}, S(g_{\text{Id}}(\bar{m},\tilde{y}^n))= g_{\text{Au}}(\bar{m},\tilde{y}^n)) \Big) \\
	&\overset{(b)}{\leq}  \sum_{m}\frac{\Big\lceil\frac{ \PP(M(g_{\text{Id}}(\bar{m},\tilde{y}^n))=m)\cdot2^{n(I(X;U,V)+2\delta_{\epsilon})}}{|\mathcal{S}|}\Big\rceil}{2^{n(I(X;U,V)+2\delta_{\epsilon})}}\\
	&\leq \sum_{m} \Big(\frac{ \PP(M(g_{\text{Id}}(\bar{m},\tilde{y}^n))=m)\cdot2^{n(I(X;U,V)+2\delta_{\epsilon})}}{|\mathcal{S}|}+1\Big)\cdot \frac{1}{2^{n(I(X;U,V)+2\delta_{\epsilon})}}\\
	&\overset{(c)}{=} 2^{-n(I(V;Y|U)-I(V;Z|U)-\delta_{\epsilon})}  + 2^{-n(I(V;Y)-R_I-3\delta_{\epsilon})} \overset{(d)}{\leq} 2^{-n(I(V;Y|U)-I(V;Z|U)-\delta_{\epsilon}')},
	\end{align*}
		where  $(a)$ follows since $\bar{M}^{\setminus g_{\text{Id}}(\bar{m},\tilde{y}^n)}$ is independent of $(M(g_{\text{Id}}(\bar{m},\tilde{y}^n)),S(g_{\text{Id}}(\bar{m},\tilde{y}^n)))$, $(b)$ follows from the uniform bin and subbin index assignment in the achievable scheme and from the bound in $\eqref{eq:joint_prob}$, $(c)$ follows from the code construction where $|\mathcal{S}|=2^{n(I(V;Y|U)-I(V;Z|U)-\delta_{\epsilon})}$ and $|\mathcal{M}|=|\mathcal{M}_1||\mathcal{M}_2|=2^{n(I(X;V|Y) + R_i+ 5\delta_{\epsilon})}$, and $(d)$ follows from the constraint $R_I < I(U;Y)$ derived for the identification rate  which together with the Markov chain $U-V-Y$ makes $I(V;Y)-R_I\geq I(V;Y|U)-I(V;Z|U)$.
	
	That is, we have
	\begin{align*}
	&\frac{1}{n}\log\frac{1}{\mbox{mFAP}} \geq I(V;Y|U)-I(V;Z|U)-\delta_{\epsilon}' \geq E-\delta_{\epsilon}',
	\end{align*}
	if $E \leq I(V;Y|U)-I(V;Z|U)$.
	
	\textbf{Converse}: We provide a converse proof for the mFAP exponent. Set $U_i \triangleq (W,M(W),Y_{i+1}^n,Z^{i-1})$ and $V_i \triangleq (W,M(W),S(W),Y_{i+1}^n,Z^{i-1})$ which satisfy $U_i-V_i-X_i(W)-(Y_i,Z_i)$ for all $i=1,\ldots,n$.
	
	Let us define the set of secret key messages that can be reconstructed from $\bar{m}$, i.e., $\mathcal{C}(\bar{m},w) = \{s(w): \text{there exists a}\ y^n\in \mathcal{Y}^n\ \text{s.t.}\ g_{\text{Au}}^{(n)}(\bar{m},y^n)=s(w)\}$. Also, let $C(\cdot)$ be a function of $s(w)$ and $\bar{m}$, where $C(s(w),\bar{m}) = 1$ for $s(w) \in \mathcal{C}(\bar{m},w)$, and $0$ otherwise. We have that
	\begin{align*}
	\delta_{n} &\geq \PP((\hat{S},\hat{W})\neq (S(W),W)) \\
	&\geq \PP(\hat{S}\neq S(W),\hat{W}=W)\\
	&\geq (1-\delta_n)\PP(\hat{S}\neq S(W)|\hat{W}=W),
	\end{align*}
	where the last inequality follows from $\PP(\hat{W}=W) \geq 1-\delta_n$.
	Thus, $\delta_{n}'\triangleq\frac{\delta_{n}}{1-\delta_{n}} \geq \PP(\hat{S}\neq S(W)|\hat{W}=W) \geq \sum_{\bar{m},w}\PP(\bar{M}=\bar{m},W=w,S(W) \notin \mathcal{C}(\bar{m},w))=\PP(C=0)$.
	
	Now consider the following bound.
	\begin{align}
	\text{mFAP} &= \max_{\tilde{y}^n(\bar{M},Z^n)\in \mathcal{Y}^n} \PP(g_{\text{Au}}^{(n)}(\bar{M},\tilde{y}^n) =S(g_{\text{Id}}^{(n)}(\bar{M},\tilde{y}^n))) \nonumber \\
	&= \max_{\tilde{y}^n} \sum_{\bar{m},z^n,w}\PP(\bar{M}=\bar{m},Z^n=z^n,W=w,g_{\text{Au}}^{(n)}(\bar{m},\tilde{y}^n) =S(g_{\text{Id}}^{(n)}(\bar{m},\tilde{y}^n))) \nonumber \\
	&\overset{(a)}{\geq} \sum_{\bar{m},z^n,w}\PP(\bar{M}=\bar{m},Z^n=z^n,W=w,S(w)=\tilde{s}(\bar{m},z^n,w) ) \nonumber \\
	&\overset{(b)}{=}\sum_{\bar{m},z^n,w}p(\bar{m},z^n,w)\max_{s(w) \in \mathcal{C}(\bar{m},w)}p(s(w)|\bar{m},z^n,w)\nonumber\\
	&\geq \sum_{\bar{m},z^n,w}p(\bar{m},z^n,w)\max_{s(w) \in \mathcal{C}(\bar{m},w)}p(s(w),C=1|\bar{m},z^n,w)\nonumber\\
	&\geq \sum_{\bar{m},z^n,w}p(\bar{m},z^n,w)p(C=1|\bar{m},z^n,w)\cdot  \max_{s(w) \in \mathcal{C}(\bar{m},w)}p(s(w)|\bar{m},z^n,w, C=1), \label{eq:FAP_converse}
	\end{align}
	where in $(a)$, the adversary who knows $\bar{m}$ and $z^n$ may choose $\tilde{y}^n$ that results in $g_{\text{Id}}^{(n)}(\bar{m},\tilde{y}^n)=w$, and the corresponding MAP estimate of $s(w)$, i.e.,
	\begin{align}
	g_{\text{Au}}^{(n)}(\bar{m},\tilde{y}^n) \triangleq\tilde{s}(\bar{m},z^n,w) = \arg\max_{s(w) \in \mathcal{C}(\bar{m},w)}p(s(w)|\bar{m},z^n,w), \label{eq:MAP_strategy_converse}
	\end{align}
	and $(b)$ follows from $\eqref{eq:MAP_strategy_converse}$.
	
	Then for any achievable $E$, it follows that
	\begin{align*}
	&n(E-\delta_n) \leq \log\Big(\frac{1}{\text{mFAP}}\Big)\\ 
	&\overset{(a)}{\leq} -\log \big(\PP(C=1)\big) -\log\big(\sum_{\bar{m},z^n,w}p(\bar{m},z^n,w|C=1)\cdot \max_{s(w)\in \mathcal{C}(\bar{m},w)}p(s(w)|\bar{m},z^n,w,C=1)\big) \\
	&\overset{(b)}{\leq} -\log(1-\delta_n') -\sum_{\bar{m},z^n,w}p(\bar{m},z^n,w|C=1)\cdot \log\big(\max_{s(w)\in \mathcal{C}(\bar{m},w)}p(s(w)|\bar{m},z^n,w,C=1)\big) \\
	&\leq -\log(1-\delta_n') -\sum_{\bar{m},z^n,w}\Big( p(\bar{m},z^n,w|C=1)\cdot  \\ &\qquad \sum_{s(w)\in \mathcal{C}(\bar{m},w)}p(s(w)|\bar{m},z^n,w,C=1)\cdot \log(p(s(w)|\bar{m},z^n,w,C=1)) \Big)\\
	&= -\log(1-\delta_n')+ H(S(W)|\bar{M},Z^n,W,C=1),
	\end{align*}
	where $(a)$ follows from $\eqref{eq:FAP_converse}$ and $(b)$ follows from $\PP(C=1)\geq 1-\delta_n'$ and Jensen's inequality
	\cite{Williams}.
	
	Continuing the chain of inequalities with the fact that 
	\[(1-\delta_n')H(S(W)|\bar{M},Z^n,W,C=1)\leq \PP(C=1) H(S(W)|\bar{M},Z^n,W,C=1) \leq H(S(W)|\bar{M},Z^n,W),\] we get
	\begin{align*}
	(1-\delta_n')\cdot[n(E-\delta_n)+\log(1-\delta_n')]
	&\leq H(S(W)|\bar{M},Z^n,W)\\
	&\overset{(a)}{\leq} \sum_{i=1}^n I(V_i;Y_i|U_i)-I(V_i;Z_i|U_i)+n\epsilon_n,
	\end{align*}
	where $(a)$ follows from the steps \eqref{eq:Econverse_start} to \eqref{eq:Econverse_end}.

	The proof ends with the standard steps for single letterization using a time-sharing random variable and letting $\delta_n, \epsilon_n \rightarrow 0$ as $n\rightarrow \infty$.
\end{IEEEproof}

\section{Conclusion}
We studied two related problems of secret key-based identification and authentication under a privacy constraint on the enrolled source data. An adversary is assumed to have access to the stored database of helping data and the ``online" side information correlated with the user's data. First, we considered the case where the adversary is passive and characterized the optimal tradeoff region of the identification rate, compression rate, leakage rate, and secret key rate for discrete memoryless sources. Then we considered a variant of the problem where  the adversary is active in the sense that it tries to deceive the identification/authentication using its own data. In this problem, we characterized the optimal tradeoff between the identification rate, compression rate, leakage rate, and mFAP exponent. Both results are derived based on the same achievability scheme involving layered random binning and rate allocation technique applied on the first layered codeword. They shed light on whether one should aim to design the secret key-based identification/authentication system to achieve the highest secret key rate as the secret key here is not for encryption but only for authentication purpose. It turned out that the maximum secret key rate in the first problem is equivalent to the maximum achievable mFAP exponent in the second one, revealing a close connection between security of identification/authentication system and the maximum achievable secret key rate. 

\appendices
\section{Proof of Lemma \ref{lemma:1}} \label{appendix:Lemma1}
Let $T$ be a binary random variable taking value $0$ if $(U^n(J(w)),Z^n) \in \mathcal{T}_{\epsilon}^{(n)}$, and $1$ otherwise. Since $(X^n(w),U^n(J(w)),V^n(J(w),K(w)),Y^n,Z^n) \in \mathcal{T}_{\epsilon}^{(n)}$ with high probability, we have $\PP(T=1) \leq \delta_{\epsilon}$. It follows that
\begin{align*}
H(Z^n|J(w)) &\overset{(a)}{\leq} H(Z^n,T|J(w),U^n)\\
&\leq H(Z^n|U^n,T) + H(T)\\
&= \PP(T=0) H(Z^n|U^n,T=0) + \PP(T=1) H(Z^n|U^n,T=1) + H(T)\\
&\overset{(b)}{\leq}H(Z^n|U^n,T=0) +\delta_{\epsilon} H(Z^n)+ h(\delta_{\epsilon}) \\
&\leq H(Z^n|U^n,T=0) + n\delta_{\epsilon} \log|\mathcal{Z}| + h(\delta_{\epsilon})\\
&= \sum_{u^n \in \mathcal{T}_{\epsilon}^{(n)}} p(u^n|T=0) H(Z^n|U^n=u^n,T=0)  + n\delta_{\epsilon} \log|\mathcal{Z}| + h(\delta_{\epsilon})\\
&\overset{(c)}{\leq} \sum_{u^n \in \mathcal{T}_{\epsilon}^{(n)}} p(u^n|T=0) \log|\mathcal{T}_{\epsilon}^{(n)}(Z|u^n)| + n\delta_{\epsilon} \log|\mathcal{Z}|  + h(\delta_{\epsilon})\leq n(H(Z|U)+\delta_{\epsilon}'),
\end{align*}
where $(a)$ follows from the fact that given the codebook, $U^n$ is a function of $J(w)$, $(b)$ follows from $\PP(T=1) \leq \delta_{\epsilon}$ where $h(\cdot)$ is the binary entropy function, and $(c)$ follows from the property of jointly typical set \cite{ElGamalKim} with $\delta_{\epsilon}, \delta_{\epsilon}' \rightarrow 0$ as $\epsilon \rightarrow 0$, and $\epsilon \rightarrow 0$ as $n \rightarrow \infty$.  
%
%
\section{Proof of Markov Chains in Converse of Theorem \ref{theorem:region_keyrate}} \label{appendix:markovchain}
We prove the Markov chains used in the converse proof of Theorem \ref{theorem:region_keyrate} based on the fact that, conditioned on $W=w$, the joint PMF of $(X^n(w),\bar{M},S(w),Y^n,Z^n)$ is given by
\begin{align*}
&\PP(X^n(w)=x^n,\bar{M}=\bar{m},S(w)=s,Y^n=y^n,Z^n=z^n|W=w) \\
&= \PP(X^n(w)=x^n,M(w)=m_w,S(w)=s|W=w) \PP(\bar{M}^{\setminus w}=\bar{m}^{\setminus w}|W=w)\cdot \\&\qquad \PP(Y^n=y^n,Z^n=z^n|X^n(w)=x^n).
\end{align*}

(I) $Y^n-M(w)-\bar{M}^{\setminus w}$ 
\begin{IEEEproof}
	Conditioned on $W=w$, we write the joint PMF of $(Y^n,\bar{M})$ as
		\begin{align*}
		&\PP(Y^n=y^n,\bar{M}=\bar{m}|W=w) \\
		&= \sum_{x^n}\PP(X^n(w)=x^n,M(w)=m_w|W=w)\PP(\bar{M}^{\setminus w}=\bar{m}^{\setminus w}|W=w)\PP(Y^n=y^n|X^n(w)=x^n)\\
		&=\PP(M(w)=m_w,Y^n=y^n|W=w)\PP(\bar{M}^{\setminus w}=\bar{m}^{\setminus w}|W=w)
		\end{align*}
		which implies that $Y^n-M(w)-\bar{M}^{\setminus w}$ forms a Markov chain.
\end{IEEEproof}

(II) $(\bar{M},S(w))-X^n(w)-(Y^n,Z^n)$
\begin{IEEEproof}
	Conditioned on $W=w$, we write the joint PMF of $(X^n(w),\bar{M},S(w),Y^n,Z^n)$ as
	\begin{align*}
	&\PP(X^n(w)=x^n,\bar{M}=\bar{m},S(w)=s,Y^n=y^n,Z^n=z^n|W=w) \\
	&= \PP(X^n(w)=x^n,M(w)=m_w,S(w)=s|W=w) \PP(\bar{M}^{\setminus w}=\bar{m}^{\setminus w}|W=w)\cdot \\&\qquad \PP(Y^n=y^n,Z^n=z^n|X^n(w)=x^n)\\
	&=\PP(X^n(w)=x^n,\bar{M}=\bar{m},S(w)=s|W=w) \PP(Y^n=y^n,Z^n=z^n|X^n(w)=x^n)
	\end{align*}
	which implies the Markov chain $(\bar{M},S(w))-X^n(w)-(Y^n,Z^n)$.
\end{IEEEproof}

(III) $(X^n(w),Z^n)-Y^n-\bar{M}^{\setminus w}$
\begin{IEEEproof}
Conditioned on $W=w$, we write the joint PMF of $(X^n(w),\bar{M}^{\setminus w},Y^n,Z^n)$ as
	\begin{align*}
	&\PP(X^n(w)=x^n,\bar{M}^{\setminus w}=\bar{m}^{\setminus w},Y^n=y^n,Z^n=z^n|W=w) \\
	&= \PP(X^n(w)=x^n,Y^n=y^n,Z^n=z^n|W=w) \PP(\bar{M}^{\setminus w}=\bar{m}^{\setminus w}|W=w)
	\end{align*}
	which implies the Markov chain $(X^n(w),Z^n)-Y^n-\bar{M}^{\setminus w}$.
\end{IEEEproof}

(IV) $(X^n(w),Y^n,Z^n,S(w))-M(w)-\bar{M}^{\setminus w}$
\begin{IEEEproof}
	Conditioned on $W=w$, we write the joint PMF of $(X^n(w),\bar{M},S(w),Y^n,Z^n)$ as
	\begin{align*}
	&\PP(X^n(w)=x^n,\bar{M}=\bar{m},S(w)=s,Y^n=y^n,Z^n=z^n|W=w) \\
	&= \PP(X^n(w)=x^n,M(w)=m_w,S(w)=s|W=w) \PP(\bar{M}^{\setminus w}=\bar{m}^{\setminus w}|W=w)\cdot \\&\qquad \PP(Y^n=y^n,Z^n=z^n|X^n(w)=x^n)\\
	&=\PP(X^n(w)=x^n,M(w)=m_w,S(w)=s,Y^n=y^n,Z^n=z^n|W=w) \PP(\bar{M}^{\setminus w}=\bar{m}^{\setminus w}|W=w)\
	\end{align*}
		which implies the Markov chain $(X^n(w),Y^n,Z^n,S(w))-M(w)-\bar{M}^{\setminus w}$.
\end{IEEEproof}
\section{Cardinality Bounds of The Sets $\mathcal{U}$ and $\mathcal{V}$ in Theorem \ref{theorem:region_exponent}} \label{appendix:cardinality}
Consider the expression of $\mathcal{R}_{1}$ in Theorem \ref{theorem:region_keyrate}:
\begin{align*}
R_I &\leq I(Y;U) \\
R_C &\geq R_I + I(X;V|Y), \\
L & \geq I(X;V,Y) - I(X;Y|U)+I(X;Z|U),\\
R_S &\leq I(V;Y|U)-I(V;Z|U),
\end{align*}
for some $U \in \mathcal{U}$, $V \in \mathcal{V}$ such that $U-V-X-(Y,Z)$ forms a Markov chain.

We can rewrite some mutual information terms in the expression above as
\begin{align*}
R_I &\leq H(Y)-H(Y|U) \\
R_C &\geq R_I+ H(X|Y)-H(X,Y|V)+H(Y|V),\\
L &\geq H(X)-H(X,Y|V)+H(Y|V) -H(Y|U)+ H(Y|X)+ H(Z|U)-H(Z|X),\\
R_S &\leq H(Y|U)-H(Y|V)-H(Z|U)+H(Z|V).
\end{align*}

We will show that the random variables $U$ and $V$ may be replaced by new ones, satisfying $|\mathcal{U}| \leq  |\mathcal{X}|+4$, $|\mathcal{V}| \leq  (|\mathcal{X}|+4)(|\mathcal{X}|+2)$, and preserving the terms $H(X,Y|V),H(Y|V),H(Z|V)$, $H(Y|U)$, and $H(Z|U)$.

First, we bound the cardinality of the set $\mathcal{U}$.
Let us define the following $|\mathcal{X}|+4$ continuous functions of $p(v|u)$, $v \in \mathcal{V}$,
\begin{align*}
&f_{j}(p(v|u)) = \sum_{v \in \mathcal{V}}p(v|u)p(x|u,v),\ j=1,\ldots,|\mathcal{X}|-1, \\
&f_{|\mathcal{X}|}(p(v|u))  = H(X,Y|V,U=u)\\ & \qquad \qquad \qquad = H(X,Y,V|U=u)-H(V|U=u),\\
&f_{|\mathcal{X}|+1}(p(v|u)) =H(Y|V,U=u)\\ & \qquad \qquad \qquad = H(Y,V|U=u)-H(V|U=u),\\
&f_{|\mathcal{X}|+2}(p(v|u)) =H(Z|V,U=u)\\ & \qquad \qquad \qquad = H(Z,V|U=u)-H(V|U=u),\\
&f_{|\mathcal{X}|+3}(p(v|u)) =H(Y|U=u),\\
&f_{|\mathcal{X}|+4}(p(v|u)) =H(Z|U=u).
\end{align*}
The corresponding averages are
\begin{align*}
& \sum_{u \in \mathcal{U}}p(u)f_{j}(p(v|u))=P_{X}(x),\ j=1,\ldots,|\mathcal{X}|-1, \\
& \sum_{u \in \mathcal{U}}p(u) f_{|\mathcal{X}|}(p(v|u))= H(X,Y,V|U)-H(V|U), \\
& \sum_{u \in \mathcal{U}}p(u) f_{|\mathcal{X}|+1}(p(v|u))=  H(Y,V|U)-H(V|U), \\
& \sum_{u \in \mathcal{U}}p(u) f_{|\mathcal{X}|+2}(p(v|u))=  H(Z,V|U)-H(V|U), \\
& \sum_{u \in \mathcal{U}}p(u) f_{|\mathcal{X}|+3}(p(v|u))= H(Y|U),\\
& \sum_{u \in \mathcal{U}}p(u) f_{|\mathcal{X}|+4}(p(v|u))= H(Z|U).
\end{align*}
According to the support lemma \cite{CsiszarBook}, we can deduce that there exists a new random variable $U'$ jointly distributed with $(X,Y,Z,V)$ whose alphabet size is $|\mathcal{U}'|= |\mathcal{X}|+4$, and numbers $\alpha_{i} \geq 0$ with $\sum_{i=1}^{|\mathcal{X}|+4}\alpha_{i} =1$ that satisfy
\begin{align*}
&\sum_{i=1}^{|\mathcal{X}|+4}\alpha_{i} f_{j}(P_{V|U'}(v|i)) = P_{X}(x),\ j=1,\ldots,|\mathcal{X}|-1, \\
&\sum_{i=1}^{|\mathcal{X}|+4}\alpha_{i}f_{|\mathcal{X}|}(P_{V|U'}(v|i)) = H(X,Y,V|U')-H(V|U'),\\
&\sum_{i=1}^{|\mathcal{X}|+4}\alpha_{i}f_{|\mathcal{X}|+1}(P_{V|U'}(v|i)) = H(Y,V|U')-H(V|U'),\\
&\sum_{i=1}^{|\mathcal{X}|+4}\alpha_{i}f_{|\mathcal{X}|+2}(P_{V|U'}(v|i)) = H(Z,V|U')-H(V|U'),\\
&\sum_{i=1}^{|\mathcal{X}|+4}\alpha_{i}f_{|\mathcal{X}|+3}(P_{V|U'}(v|i)) = H(Y|U'),\\
&\sum_{i=1}^{|\mathcal{X}|+4}\alpha_{i}f_{|\mathcal{X}|+4}(P_{V|U'}(v|i)) = H(Z|U').
\end{align*}
Note that we have
\begin{align*}
&H(X,Y,V|U')-H(V|U') \\
&= H(X,Y,V|U)-H(V|U)\\
& \overset{(a)}{=} H(X,Y|V),
\end{align*}
where $(a)$ follows from the Markov chain $U-V-X-(Y,Z)$.
Similarly, from the Markov chain $U-V-X-(Y,Z)$, we have that $H(Y,V|U')-H(V|U')=H(Y,V|U)-H(V|U)=H(Y|V)$, and $H(Z,V|U')-H(V|U')=H(Z,V|U)-H(V|U)=H(Z|V)$.
Since $P_{X}(x)$ is preserved, $P_{X,Y,Z}(x,y,z)$ is also preserved.  Thus, $H(X|Y), H(Y|X), H(Z|X)$ are preserved.

Next we bound the cardinality of the set $\mathcal{V}$.
For each $u' \in \mathcal{U}'$, we define the following $|\mathcal{X}|+2$ continuous functions of $p(x|u',v)$,\ $x \in \mathcal{X}$,
\begin{align*}
&f_{j}(p(x|u',v)) = p(x|u',v),\ j=1,\ldots,|\mathcal{X}|-1, \\
&f_{|\mathcal{X}|}(p(x|u',v)) = H(X,Y|U'=u',V=v),\\
&f_{|\mathcal{X}|+1}(p(x|u',v)) = H(Y|U'=u',V=v),\\
&f_{|\mathcal{X}|+2}(p(x|u',v)) = H(Z|U'=u',V=v).
\end{align*}
Similarly to the previous part in bounding $|\mathcal{U}|$, there exists a new random variable $V'|\{U'=u'\} \sim p(v'|u')$ such that $|\mathcal{V}'| =  |\mathcal{X}|+2$ and $p(x|u')$, $H(X,Y|U'=u',V)$, $H(Y|U'=u',V)$, and $H(Z|U'=u',V)$ are preserved.

By setting $V'' =(V',U')$ where $\mathcal{V}'' = \mathcal{V}' \times \mathcal{U}'$, we have that $U'-V''-X-(Y,Z)$ forms a Markov chain.

Furthermore, we have the following preservations by $V''$,
\begin{align*}
&H(X,Y|V'') \\
& = H(X,Y|V',U')\\
& \overset{(a)}{=} H(X,Y|V,U')\\
& \overset{(b)}{=} H(X,Y|V,U)\\
& \overset{(c)}{=} H(X,Y|V),
\end{align*}
where $(a)$ follows from preservation by $V'$, $(b)$ follows from preservation by $U'$, and $(c)$ follows from the Markov chain $U-V-X-(Y,Z)$.
Similarly, from preservation by $U'$ and $V'$, and the Markov chain $U-V-X-(Y,Z)$, we have that $H(Y|V'') =H(Y|V',U')=H(Y|V)$ and $H(Z|V'')=H(Z|V',U')=H(Z|V)$.

Therefore, we have shown that $U \in \mathcal{U}$ and $V \in \mathcal{V}$ may be replaced by $U' \in \mathcal{U}'$ and $V'' \in \mathcal{V}''$ satisfying
\begin{align*}
|\mathcal{U}'|&= |\mathcal{X}|+4, \\
|\mathcal{V}''| &= |\mathcal{U}'||\mathcal{V}'|= (|\mathcal{X}|+4)(|\mathcal{X}|+2),
\end{align*}
and  preserving the terms $H(X,Y|V),H(Y|V),H(Z|V)$, $H(Y|U)$, and $H(Z|U)$.

\section{Proof of the Compression-leakage-key rate Region in the Binary Example} \label{appendix:proof_example}
\emph{Achievability:}  Let $V$ be an output of a BSC($\alpha$) with input $X$, where $\alpha \in [0,1/2]$. Then by setting $U=\emptyset$, it follows from the expression in Remark 2 i) that
\begin{align*}
R_C &\geq I(X;V|Y) \\
&\overset{(a)}{=}p\cdot(H(X)-H(X|V)) \\
&\overset{(b)}{=}p\cdot(1-h(\alpha)),
\end{align*}
where $(a)$ follows since $Y=e$ with probability $p$, otherwise $Y=X$, and $(b)$ follows from the choice of $V$,
\begin{align*}
L &\geq I(X;Z)+ I(X;V|Y)\\
&\overset{(a)}{=} 1-H(X|Z) + p\cdot(1-h(\alpha))\\
&\overset{(b)}{=} 1-((1-p)q +p)+ p\cdot(1-h(\alpha))\\
&= (1-q)(1-p) + p\cdot(1-h(\alpha)),
\end{align*}
where $(a)$ follows from the bound on $R_C$ and $(b)$  follows since $Z=e$ with probability $(1-p)q +p$, otherwise $Z=X$.
\begin{align*}
R_S &\leq I(Y;V|Z)\\
&\overset{(a)}{=}  I(X;V|Z)-I(X;V|Y)\\
&\overset{(b)}{=} ((1-p)q +p)\cdot I(X;V)- p\cdot(1-h(\alpha))\\
&= q (1-p)(1-h(\alpha)),
\end{align*}
where  $(a)$  follows from the Markov chain $V-X-Y-Z$ and $(b)$ follows since $Z=e$ with probability $(1-p)q +p$, otherwise $Z=X$.

\emph{Converse:}
Let $(R_C,L,R_S)$ be an achievable tuple. We now prove that there exist $\alpha \in [0,1/2]$ satisfying the inequalities shown in the achievability above. From the region specified in  Remark 2 i), we have the following bound on the compression rate $R_C$.
\begin{align*}
R_C &\geq I(X;V|Y) \\
&= p\cdot I(X;V)  \\
& = p\cdot(1-H(X|V)).
\end{align*}
Since $0 \leq H(X|V) \leq H(X)=1$, and $h(\cdot)$ is a continuous one-to-one mapping from $[0,1/2]$ to $[0,1]$, there exists $\alpha \in [0,1/2]$ s.t. $H(X|V)=h(\alpha)$, and thus $R_C \geq p\cdot(1-h(\alpha))$. The bounds on $L$ and $R_S$ readily follow from $H(X|V)=h(\alpha)$.


\end{document}